\documentclass[journal,twoside,web]{ieeecolor}
\usepackage{generic}
\usepackage{cite}
\usepackage{amsmath,amssymb,amsfonts}
\usepackage{algorithmic}
\usepackage{graphicx}
\usepackage{algorithm,algorithmic}
\usepackage{hyperref}
\usepackage{multirow}
\usepackage{comment}
\usepackage[]{acronym}
\usepackage{colortbl}
\definecolor{tablegray}{gray}{0.8}

\acrodef{DL}{Deep Learning}
\acrodef{US}{ultrasound}
\acrodef{CNN}{Convolutional Neural Network}
\acrodef{CNNs}{Convolutional Neural Networks}
\acrodef{ROI}{Region of Interest}
\acrodef{SSL}{Semi-Supervised Learning}
\acrodef{GT}{Ground Truth}
\acrodef{CR}{Consistency Regularization}
\acrodef{EMA}{Exponential Moving Average}
\acrodef{MAC}{Mutual Agreement Consistency}
\acrodef{KL}{Kullback-Leibler divergence}
\acrodef{MIG}{Mutual Information Gap}
\acrodef{CE}{Cross-Entropy}
\acrodef{Dice}{Dice based coefficient}
\acrodef{DSC}{Dice Similarity Coefficient}
\acrodef{ASD}{Average Surface Distance}
\acrodef{HD95}{95\% Hausdorff Distance}
\acrodef{SOTA}{state-of-the-art}
\acrodef{PS}{Pubic Symphysis}
\acrodef{FH}{Fetal Head}
\acrodef{NCA}{Neural Cellular Automata}
\acrodef{ViT}{Vision Transformer}
\acrodef{POCUS}{Point-of-Care Ultrasound}
\acrodef{POC}{point-of-care}
\acrodef{CCA}{Common Carotid Artery}
\acrodef{ReLU}{Rectified Linear Unit}
\acrodef{GELU}{Gaussian Error Linear Unit}
\acrodef{FiLM}{Feature-wise Linear Modulation}
\acrodef{AI}{Artificial Intelligence}
\acrodef{MLP}{multilayer perceptron}

\hypersetup{hidelinks=true}
\usepackage{textcomp}
\def\BibTeX{{\rm B\kern-.05em{\sc i\kern-.025em b}\kern-.08em
    T\kern-.1667em\lower.7ex\hbox{E}\kern-.125emX}}
\begin{document}
\title{Efficient Ultrasound Image Segmentation with Token-Conditioned Neural Cellular Automata}

\author{Fangyijie Wang*, Tanya Akumu, Zi Ye, Gu\'enol\'e Silvestre, Karim Lekadir, Kathleen M. Curran*
\thanks{Fangyijie Wang and Kathleen M. Curran are the corresponding authors. Fangyijie Wang, Kathleen M. Curran are affiliated with the School of Medicine, University College Dublin, Ireland and the Research Ireland Centre for Research Training in Machine Learning (e-mail: fangyijie.wang@ucdconnect.ie, kathleen.curran@ucd.ie).}
\thanks{Gu\'enol\'e Silvestre is affiliated with the School of Computer Science, University College Dublin, Dublin, Ireland and the Research Ireland Centre for Research Training in Machine Learning.}
\thanks{Zi Ye is affiliated with the Department of Computer Science, Maynooth University, Maynooth, Ireland.}
\thanks{Tanya Akumu, Karim Lekadir are affiliated with the Departament de Matem\`atiques i Inform\`atica, Universitat de Barcelona, Barcelona, Spain. Karim Lekadir is also with Instituci\'o Catalana de Recerca i Estudis Avan\c{c}ats (ICREA).}
}

\maketitle

\begin{abstract}
\ac{POCUS} plays an important role in bedside diagnosis and clinical decision-making, particularly in resource-constrained settings. Recent deep learning methods have substantially improved ultrasound image segmentation, enabling accurate diagnosis and biometric estimation. However, their computational cost limits deployment on portable and low-resource devices. To address this challenge, we propose LiteAdaNCA-Net (LANCANet), a lightweight ultrasound segmentation framework that incorporates token-conditioned Neural Cellular Automata (NCA) adapters for iterative feature refinement. Specifically, structure-aware tokens guide local NCA refinement via Token FiLM, enabling boundary-aware feature refinement with minimal computational cost. We evaluate LANCANet on HC18, CCA, and PSFHS, and assess robustness on two independent African fetal head datasets collected from multiple clinical centers. Experimental results demonstrate that LANCANet achieves competitive or superior performance to recent lightweight CNN- and transformer-based methods. On HC18 and CCA, LANCANet achieves the highest Dice Similarity Coefficient (DSC) of 96.62\% and 92.86\%, respectively. On PSFHS, it achieves the best performance on the challenging pubic symphysis structure while maintaining competitive fetal head segmentation accuracy. Furthermore, despite being trained from scratch, LANCANet maintains competitive performance on the external KEN-FH and AFR-FH datasets under substantial domain shifts. These results show that token-conditioned NCA refinement improves segmentation accuracy and boundary localization while maintaining computational efficiency for resource-constrained clinical deployment. Our code is on \href{https://anonymous.4open.science/r/LANCAN-21A0/README.md}{GitHub}.
\end{abstract}


\section{Introduction}
\label{sec:introduction}

Ultrasound image segmentation is critical for fetal monitoring, obstetric assessment, and \acf{POC} clinical decision-making~\cite{Salomon:2011,Pietsch:2021}. Because of its portability, real-time imaging capability, and cost-effectiveness, ultrasound has become one of the most widely used imaging modalities in clinical practice~\cite{Lee:2020,Self:2022}. However, accurate segmentation of ultrasound images remains challenging due to imaging artifacts such as speckle noise, acoustic shadowing, signal dropout, low contrast, and variability across devices and operators~\cite{Noble:2006,Zhou:2020}. In addition, many clinically relevant anatomical structures are small, ambiguous, or poorly delineated, making reliable boundary localization particularly difficult~\cite{Noble:2006,Zhou:2020,Boumeridja:2025,Ferreira:2025}. These challenges highlight the need for segmentation methods that are not only accurate and robust, but also computationally efficient for real-world and resource-constrained clinical deployment.

\ac{CNNs}, particularly U-shaped networks~\cite{Ronneberger:2015}, have become the dominant paradigm for medical image segmentation and have demonstrated strong performance in ultrasound analysis~\cite{Liu:2020}. However, \ac{CNNs} primarily rely on local receptive fields, which can limit their ability to capture long-range contextual information~\cite{Dosovitskiy:2020_b}. Therefore, they often fail to accurately delineate weak boundaries, irregular shapes, or small anatomical structures commonly observed in ultrasound images~\cite{Bian:2025}. More recently, transformer-based methods~\cite{Dosovitskiy:2020_b} have shown improved global context modeling capability, but their high computational complexity and large parameter requirements often restrict deployment in resource-constrained or \ac{POC} environments~\cite{Valanarasu:2022}. To address these limitations, several lightweight segmentation networks have been proposed for efficient ultrasound analysis~\cite{Zhou:2022,Valanarasu:2022,Tang:2024}. Nevertheless, reducing model complexity often weakens feature representation and segmentation robustness. This limitation is prominent in ultrasound imaging, where noise, low contrast, and ambiguous anatomical boundaries already make accurate segmentation difficult. Consequently, existing approaches still face a fundamental trade-off between segmentation accuracy, computational efficiency, and deployment feasibility in real-world clinical settings.

\ac{NCA} has recently emerged as a promising approach for efficient image segmentation through iterative local interaction modeling~\cite{Gilpin:2019}. Unlike conventional deep networks that rely on increasingly complex architectures, \ac{NCA} updates feature representations through repeated interactions between neighboring cells, enabling adaptive local refinement with relatively low computational cost. Gilpin et al.~\cite{Gilpin:2019} demonstrated that cellular automata can be represented using \ac{CNNs}, while Mordvintsev et al.~\cite{Mordvintsev:2020} further introduced neural cellular automata for modeling biological morphogenesis. Building upon these ideas, recent studies have explored \ac{NCA}-based methods in medical image analysis tasks~\cite{Kalkhof:2023,Kalkhof:2023b,Mittal:2025,Krumb:2025,Yang:2026,Deutges:2026}. However, many existing approaches employ \ac{NCA} as the primary segmentation backbone, which can limit high-level semantic modeling, hierarchical feature representation, and global contextual understanding. These limitations are particularly important in ultrasound imaging, where robust semantic representation is critical for handling weak boundaries and challenging anatomical structures. Therefore, rather than using \ac{NCA} as a standalone encoder, it may be more effective to leverage \ac{NCA} as a lightweight refinement mechanism that complements stronger semantic feature extractors.

To address these limitations, we propose LANCANet, a lightweight ultrasound segmentation framework that integrates token-conditioned \ac{NCA} adapters for adaptive feature refinement. Inspired by~\cite{Xu:2024b}, the proposed framework incorporates \ac{NCA} modules as lightweight refinement adapters rather than replacing the main feature extraction backbone. This design preserves high-level semantic representations while enabling iterative local interactions for boundary-aware feature refinement. By leveraging token-cell interactions, LANCANet improves the modeling of fine anatomical structures and ambiguous boundaries observed in ultrasound imaging. In addition, the proposed adapter design introduces only limited computational overhead, making the framework suitable for resource-constrained deployment scenarios. Through the combination of semantic feature preservation and efficient local refinement, LANCANet aims to improve segmentation accuracy and robustness without substantially increasing model complexity.

The main contributions of this work are summarized as follows:
\begin{itemize}
    \item We propose LANCANet, a lightweight ultrasound segmentation framework that incorporates token-conditioned neural cellular automata (NCA) adapters to refine intermediate representations while preserving high-level semantic features iteratively.

    \item We introduce a token-guided NCA refinement mechanism that combines global structural context with local cellular interactions to improve boundary delineation and the segmentation of small and challenging anatomical structures with limited computational cost.

    \item We provide a comprehensive evaluation of computational efficiency by comparing parameter count, computational complexity, and CPU inference speed against representative lightweight segmentation baselines. LANCANet shows a favorable balance between accuracy and efficiency.

    \item We validate LANCANet on multiple ultrasound datasets, including external evaluation under domain shift. We also perform statistical analysis to demonstrate its robustness and generalizability across different anatomical structures and imaging settings.
\end{itemize}
To the best of our knowledge, this is the first work to integrate token-conditioned \ac{NCA} adapters for ultrasound segmentation.

\section{Related Work}

\subsection{Lightweight Segmentation Networks for Medical Image}

Deep learning has significantly advanced medical image segmentation, with \ac{CNN}~\cite{Ronneberger:2015,Isensee:2021} becoming the dominant approach due to their strong representation capability. However, their deployment in portable and \ac{POC} ultrasound systems remains challenging due to computational and latency constraints. To improve efficiency, recent studies have explored lightweight segmentation architectures by reducing model complexity while maintaining segmentation accuracy. MobileNet~\cite{Sandler:2018} adopts depthwise separable convolutions. UNeXt~\cite{Valanarasu:2022} integrates convolutional encoding and tokenized \ac{MLP} blocks to achieve efficient image segmentation with reduced computational cost. Zhou et al.~\cite{Ruan:2023} introduce group-wise attention and feature aggregation modules to achieve low computational overhead. CMUNeXt~\cite{Tang:2024} leverages large-kernel convolutions and efficient feature fusion to improve global context modeling in lightweight medical image segmentation.

More recently, lightweight transformer-based models have been introduced to overcome the limitation of \ac{CNN} models in capturing global contextual information. SegFormer~\cite{Xie:2021} extended lightweight segmentation to transformer architectures by employing hierarchical transformer encoding for efficient global feature modeling. LB-UNet~\cite{Xu:2024} introduces lightweight attention and auxiliary prediction modules to reduce computational overhead while enhancing boundary-aware segmentation. Rahman et al.~\cite{Rahman:2025} proposed MK-UNet which combines multi-kernel depth-wise convolutions and lightweight attention mechanisms to improve multi-scale feature representation. Zhong et al.~\cite{Zhong:2025} introduced a lightweight model PMFSNet that integrates simplified self-attention and multi-scale feature enhancement for efficient global and local representation learning.

Despite promising results in general medical imaging tasks, the performance of these lightweight architectures in ultrasound segmentation remains insufficiently investigated. Furthermore, balancing segmentation accuracy with computational efficiency remains difficult. Lightweight models often suffer from reduced representation capacity, which can limit robustness and boundary delineation performance. This issue is particularly challenging in ultrasound imaging because of speckle noise, low contrast, and ambiguous anatomical boundaries. Therefore, there is a need for lightweight segmentation frameworks that improve local feature refinement while preserving strong semantic representations. In this work, we address this challenge through an \ac{NCA}-based refinement mechanism for efficient ultrasound segmentation.

\subsection{Neural Cellular Automata for Iterative Feature Refinement}

Neural Cellular Automata (\ac{NCA}) extend traditional cellular automata by replacing predefined update rules with learnable neural functions to enable complex pattern formation~\cite{Gilpin:2019,Mordvintsev:2020}. Unlike conventional deep networks that progressively aggregate information by increasing network depth, \ac{NCA} updates feature states repeatedly based on neighboring cells. Thus, it allows local information propagation with low computational resource requirement. These properties make \ac{NCA} attractive for tasks requiring efficient structural modeling and boundary-aware refinement.

Recent studies have explored \ac{NCA} in medical image analysis, including classification, segmentation, image reconstruction, and representation learning~\cite{Kalkhof:2023,Kalkhof:2023b,Mittal:2025,Krumb:2025,Yang:2026,Deutges:2026}. Existing approaches employ \ac{NCA} as a standalone backbone or iterative prediction framework. However, such designs may exhibit limitations in capturing high-level semantics and global contextual information. We believe \ac{NCA} can be considered as an efficient refinement mechanism to complement semantic representations learned by modern segmentation networks. Motivated by this idea, our work integrates lightweight \ac{NCA} modules as refinement adapters to iteratively improve local feature representations while preserving semantic representations extracted by the backbone.

\section{Methodology}

\begin{figure*}
\centering
\includegraphics[width=\textwidth]{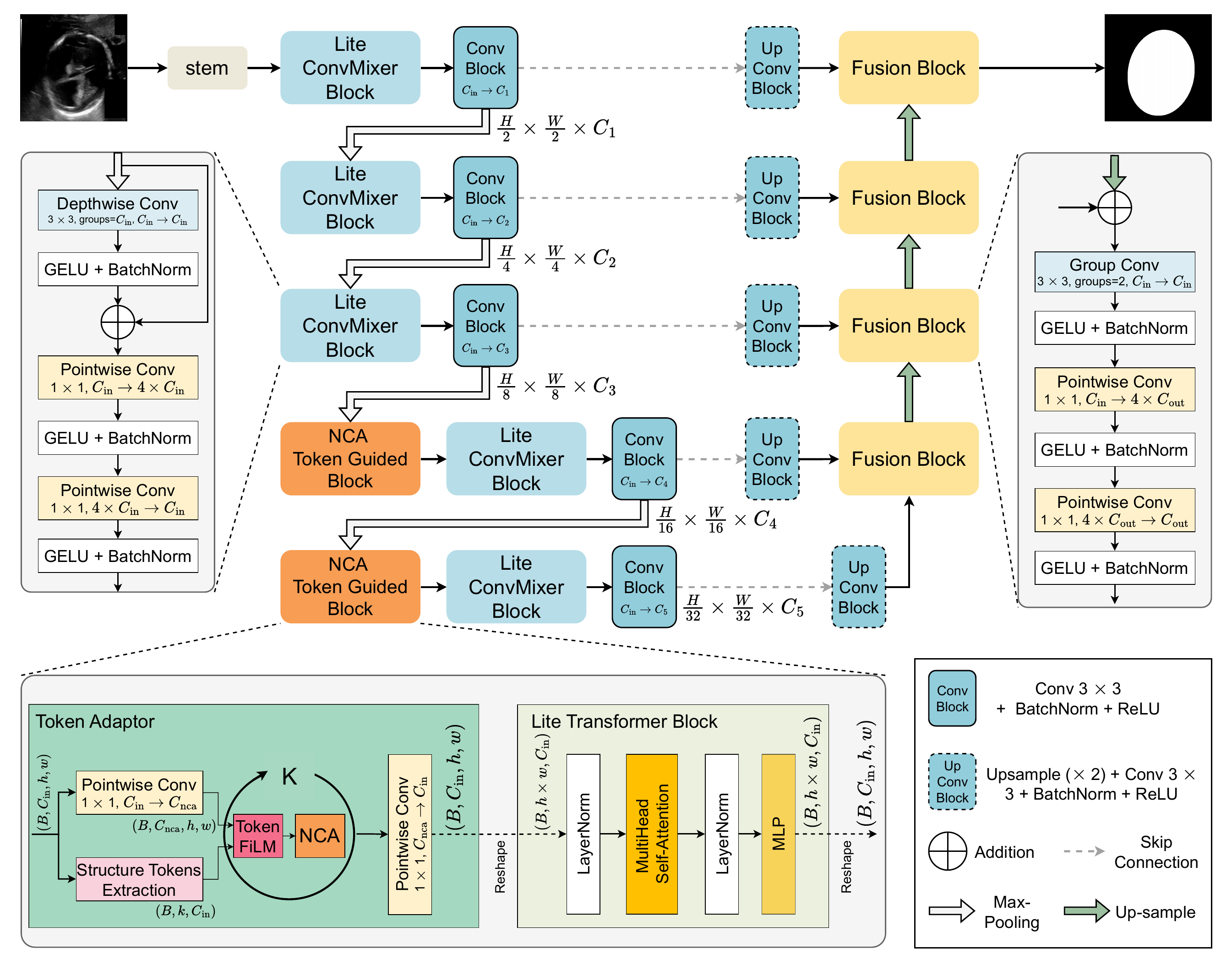}
\caption{
Overview of the proposed LANCANet architecture. Token-conditioned \ac{NCA} adapters iteratively refine feature representations using global structure-aware guidance and local cellular interactions, enabling robust and computationally efficient ultrasound image segmentation.}
\label{lancan}
\end{figure*}

\subsection{Preliminary}

\acf{NCA} models spatial data as a grid of cells whose states evolve through repeated local interactions governed by a learnable update rule~\cite{Wulff:1992,Gilpin:2019}. Given a cell-state tensor $\mathbf{S}^t \in \mathbb{R}^{h \times w \times C}$ at iteration $t$, each cell updates its state by aggregating information from its $3 \times 3$ neighborhood and applying a shared neural function:
$
\mathbf{S}^{t+1}=\mathbf{S}^t+f_u\left(f_p\left(\mathbf{S}^t\right)\right),
$
where $f_p$ denotes a local perception operator (e.g., convolutional filters) that aggregates neighborhood information, and $f_u$ is a learnable update function shared across all cells. Together, these components define the \ac{NCA} state transition.
Through iteration, \ac{NCA} learns complex global structures from simple local update rules while remaining parameter-efficient and robust.

\subsection{Encoder Stage}
The encoder stage is illustrated in Fig.~\ref{lancan}. It comprises five levels from top to bottom. The stem block extracts initial features from the input image at the highest level. The top three levels consist of a Lite ConvMixer block, a convolution block, and a down-sampling operation, whereas the bottom two levels additionally include an \ac{NCA} Token Guided block. Max pooling is used for down-sampling between levels, with a $2 \times 2$ window and a stride of 2. The stem block consists of a $3\times3$ convolution (stride 1, padding 1), followed by a batch normalization and \ac{ReLU}.

\subsubsection{Lite ConvMixer Block} The core component of the Lite ConvMixer block is depthwise separable convolution~\cite{Trockman:2023}, which has been popularized by~\cite{Tang:2023,Tang:2024}. Following their design, we use a single-layer Lite ConvMixer block, defined as:

\begin{equation}
\begin{aligned}
f_1^{\prime}&=\operatorname{BN}\left(\operatorname{GELU}\left(\operatorname{DWConv}_{3\times3}\left(f_0\right)\right)\right)+ f_{\text{in}} \\
f_1^{\prime \prime}&=\operatorname{BN}\left(\operatorname{GELU}\left(\operatorname{PWConv}_{1\times1}^{4C \rightarrow C}\left(f_1^{\prime}\right)\right)\right) \\
f_1&=\operatorname{BN}\left(\operatorname{GELU}\left(\operatorname{PWConv}_{1\times1}^{C \rightarrow 4C}\left(f_1^{\prime \prime}\right)\right)\right) \\
f_{\mathrm{out}}&=\left(\operatorname{GELU}\left(\operatorname{ReLU}\left(\operatorname{BN}\left(\mathrm{Conv}_{3\times3}^{C \rightarrow C_{\mathrm{out}}}\left(
f_1\right)\right)\right)\right)\right)
\end{aligned}
\end{equation}
where $f_1$ denotes the output feature map of layer 1 in the ConvMixer block. $\phi(\cdot)=\mathrm{BN}(\mathrm{GELU}(\cdot))$. $\mathrm{GELU}$ denotes the \ac{GELU} activation function and $\mathrm{BN}$ denotes batch normalization. $\mathrm{DWConv}$ and $\mathrm{PWConv}$ refer to depthwise and pointwise convolutions, respectively.

\subsubsection{Token Adapter} This adapter comprises three components for structure-aware refinement: {\it Structure Token Extraction}, which summarizes global structural information into compact token embeddings; {\it Token \ac{FiLM}}~\cite{Perez:2018}, which adaptively conditions the \ac{NCA} state via channel-wise modulation; and \ac{NCA}, which refines features through shared local update rules in a parameter-efficient manner (Fig.~\ref{lancan}). The refinement process is repeated for $T$ iterations. Each iteration consists of one {\it Token \ac{FiLM}} modulation step followed by one \ac{NCA} cellular update. Larger values of $T$ enable additional iterative refinement and information propagation at the cost of increased computation.

\begin{figure}
\centering
\includegraphics[width=\linewidth]{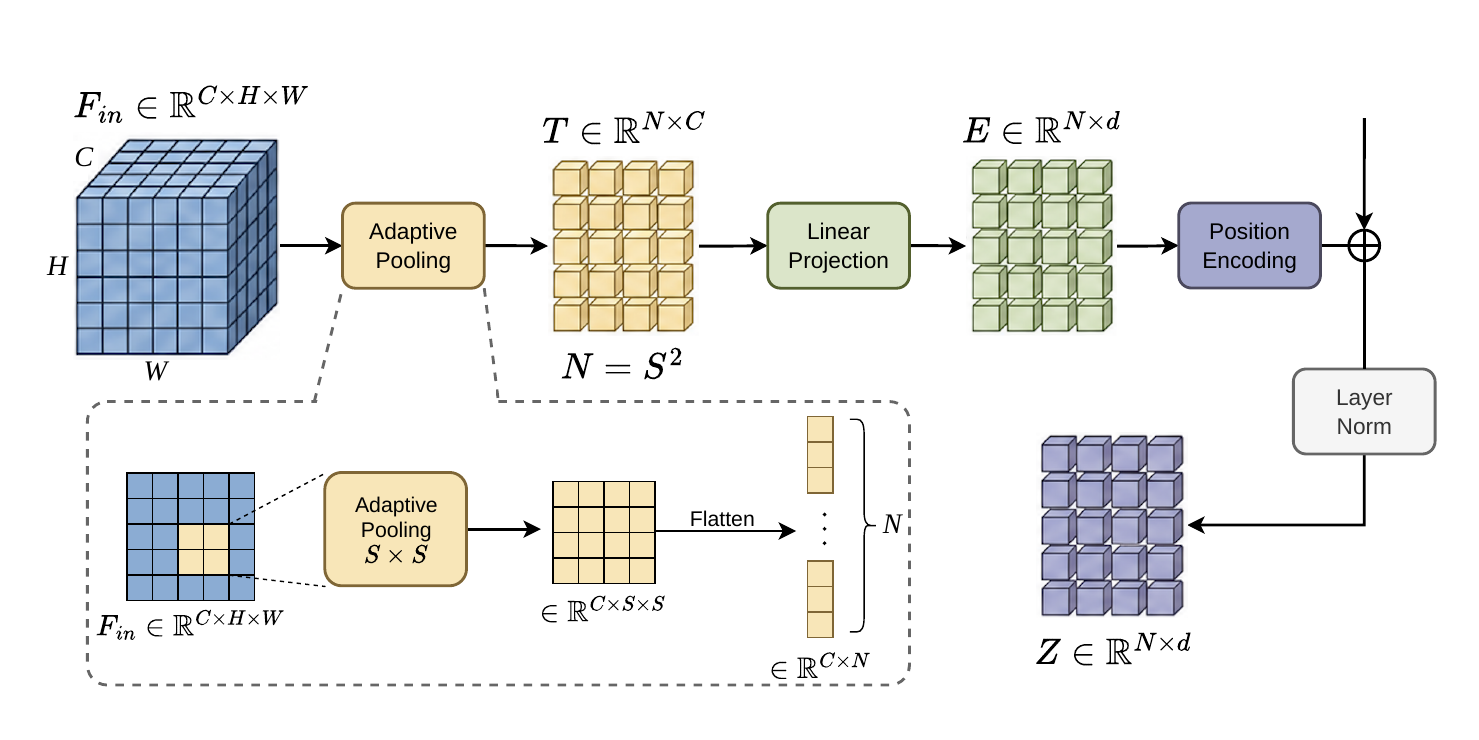}
\caption{The flow of {\it Structure Token Extraction} (Cell $\rightarrow$ Token).}
\label{struc_tokens}
\end{figure}

\textit{Structure Token Extraction (Cell $\rightarrow$ Token).} 
We first flatten the feature map $\mathbf{F}_{in} \in \mathbb{R}^{B \times C \times H \times W}$ into $N = H \times W$ spatial cells, obtaining $\mathbf{X} \in \mathbb{R}^{B \times N \times C}$. Then we extract a compact structural representation via adaptive pooling with output size $S \times S$ to get structure tokens $\mathbf{T} \in \mathbb{R}^{B \times N \times C}$, where $N = S^2$. These tokens $\mathbf{T}$ attend to the spatial features to summarize global structural information. Specifically, the tokens interact with the spatial features through scaled dot-product attention:
\begin{equation}
\mathbf{E}=\operatorname{softmax}\left(\frac{Q(\mathbf{T}) K(\mathbf{X})^{\top}}{\sqrt{d}}\right) V(\mathbf{X}),
\end{equation}
where $Q(\cdot)$, $K(\cdot)$, and $V(\cdot)$ denote linear projections and $d$ is the embedding dimension. The resulting embedded tokens $\mathbf{E} \in \mathbb{R}^{B \times N \times d}$ are further transformed by positional encoding and layer normalization to get the final structure-aware token embeddings $\mathbf{Z} \in \mathbb{R}^{B \times N \times d}$ (Fig.~\ref{struc_tokens}). Therefore, the tokens $\mathbf{Z}$ summarize global structural information and are subsequently used to guide the \ac{NCA}-based refinement.

\textit{Token \ac{FiLM} (Token $\rightarrow$ Cell Conditioning).} Inspired by Feature-wise Linear Modulation~\cite{Perez:2018}, we insert a \ac{FiLM} layer before the \ac{NCA} layer to inject global structural context into the cellular state. Specifically, the {\it Token \ac{FiLM}} module takes the structure-aware token embeddings $\mathbf{Z} \in \mathbb{R}^{B \times N \times d}$ from {\it Structure Token Extraction} as conditioning inputs to modulate the \ac{NCA} cellular state. We first aggregate the structure-aware token embeddings $\mathbf{Z} \in \mathbb{R}^{B \times N \times d}$ via mean pooling to obtain a global token summary:
\begin{equation}
\mathbf{t}=\frac{1}{N} \sum_{i=1}^N \mathbf{Z}_i, \quad \mathbf{t} \in \mathbb{R}^{B \times d}
\end{equation}
where $N$ denotes the number of structure tokens and $d$ is the token embedding dimension. This summary token is passed through a lightweight \ac{MLP} to generate channel-wise modulation parameters $\gamma, \beta \in \mathbb{R}^{B \times C}$. Then $\gamma$ and $\beta$ are applied to the \ac{NCA} state $\mathbf{S} \in \mathbb{R}^{B \times C_{nca} \times H \times W}$ via feature-wise linear modulation: $\mathbf{S}' = \mathbf{S} \odot (1+\gamma) + \beta.$ This conditioning allows the \ac{NCA} to adaptively refine local cell updates based on the global structural context encoded by the tokens.

\begin{figure}
\centering
\includegraphics[width=\linewidth]{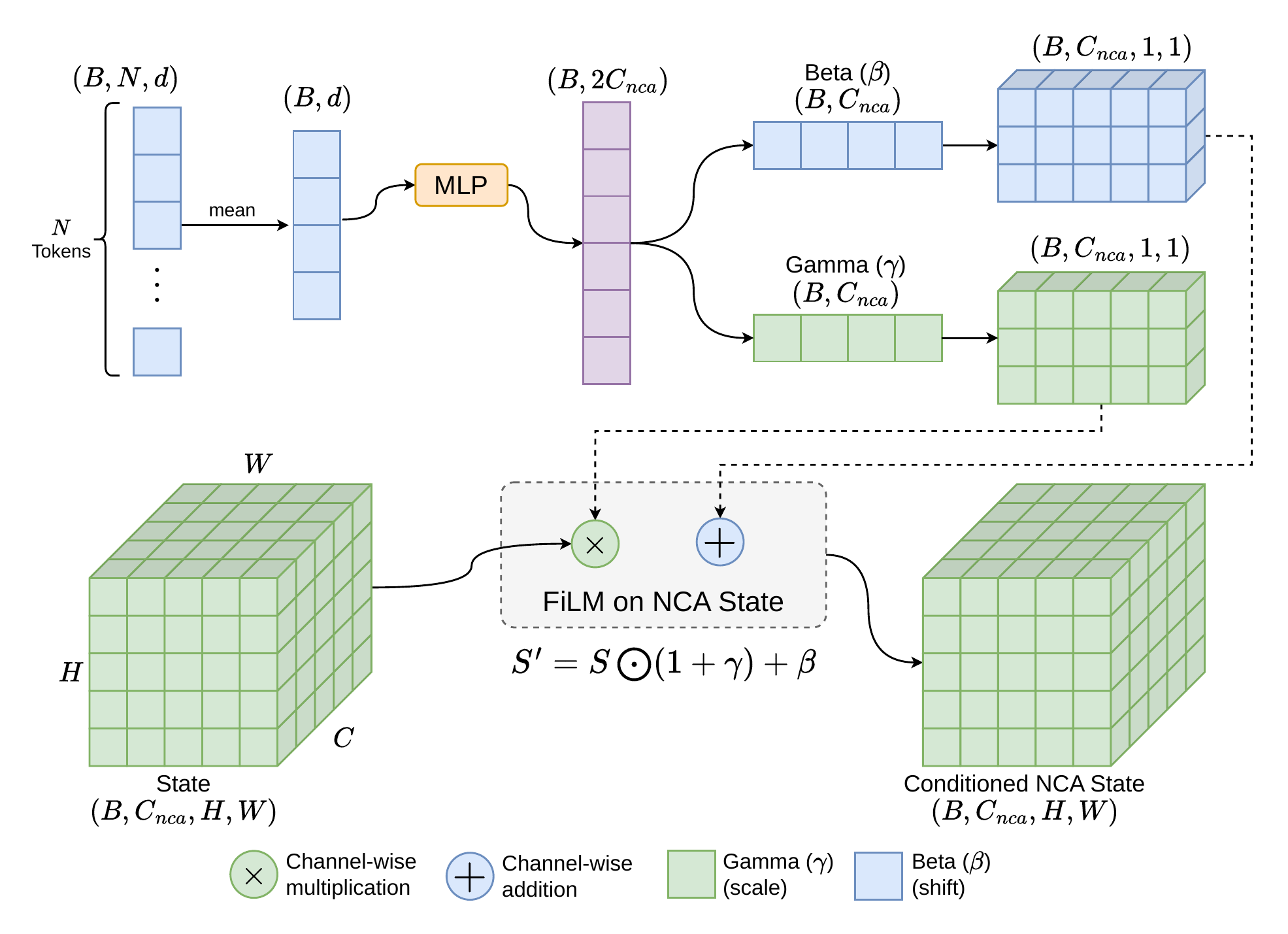}
\caption{The flow of Token \ac{FiLM} (Token $\rightarrow$ Cell Conditioning).}
\label{film}
\end{figure}

\textit{\acf{NCA} Update.} Conditioned by {\it Token \ac{FiLM}}, the \ac{NCA} dynamics adapt their local updates according to the global structural context. Our \ac{NCA} implementation follows the design of Med-NCA~\cite{Kalkhof:2023}, where each cell updates its state based on locally perceived neighborhood information and a shared, learnable update rule. Specifically, a local perception operator and a shared neural update function iteratively refine spatial features through residual, stochastic updates in a parameter-efficient manner. 

In short, the {\it Token Adapter} performs iterative refinement for $T$ iterations. In each iteration, {\it Token \ac{FiLM}} first injects global structural context into the cellular state $\mathbf{S}$, followed by an \ac{NCA} update step $f_{\mathrm{NCA}}$ that refines local spatial representations~\cite{Kalkhof:2023}. Therefore, each refinement iteration consists of one {\it Token \ac{FiLM}} conditioning operation and one \ac{NCA} cellular update. The refinement process can be expressed as:
\begin{equation}
\mathbf{S}'_{t+1}=f_{\mathrm{NCA}}\left(f_{\mathrm{TokenFiLM}}\left(\mathbf{S}'_t, \mathbf{Z}\right)\right), \quad t=1, \ldots, T
\end{equation}
where $\mathbf{Z}$ is the structure-aware token embeddings and $T$ denotes the number of refinement steps. 

\subsubsection{Lite Transformer Block} After the {\it Token Adapter}, we introduce a lightweight transformer block. This block provides global context modeling at a low computational cost by applying attention over the flattened spatial tokens while preserving the original feature map layout. Specifically, we reshape the input feature map $F \in \mathbb{R}^{B \times C_\text{in} \times H \times W}$ into a token sequence $\mathbf{X}_{\text{seq}} \in \mathbb{R}^{B \times N \times C}$ with $N = H \times W$. The token sequence $\mathbf{X}_{\text{seq}}$ is then fed into a lightweight pre-norm transformer that incorporates residual connections:
\begin{equation}
\begin{aligned}
\mathbf{X}^{\prime}&=\mathbf{X}_{\text {seq }}+\mathrm{DP}\left(\mathrm{Attn}\left(\mathrm{LN}\left(\mathbf{X}_{\text {seq }}\right)\right)\right), \\ \mathbf{X}_{\text{nca}}&=\mathbf{X}^{\prime}+\mathrm{DP}\left(\mathrm{MLP}\left(\mathrm{LN}\left(\mathbf{X}^{\prime}\right)\right)\right),
\end{aligned}
\end{equation}
where $\mathrm{LN}(\cdot)$ denotes LayerNorm, $\mathrm{Attn}(\cdot)$ is a multi-head self-attention mechanism introduced in~\cite{Vaswani:2017}, $\mathrm{MLP}(\cdot)$ is a feed-forward network with expansion ratio $r$, and $\mathrm{DP}(\cdot)$ indicates optional stochastic depth (DropPath). Finally, $\mathbf{X}_{\text{nca}}$ is reshaped back to $\mathbb{R}^{B \times C_\text{in} \times h \times w}$.

\subsection{Decoder Stage}
The decoder consists of five levels arranged from bottom to top. Each level comprises a Fusion block and an upsampling convolutional block.

\subsubsection{Upsampling Convolutional Block} This block includes an upsampling layer, a convolutional layer, a batch normalization layer, and a \ac{ReLU} activation function. The upsampling layer uses bilinear interpolation to upsample the feature maps by a factor of two. The convolutional layer has a kernel size of $3 \times 3$, with a stride of 1 and padding of 1. 

\subsubsection{Fusion Block} The fusion block leverages group convolution to reduce computational cost while preserving representational capacity. The convolution is split into two groups to independently process the encoder skip features and the upsampled decoder features. The group convolution employs a $3 \times 3$ kernel with stride 1 and padding 1, followed by two inverted bottleneck pointwise convolutions for enhanced feature fusion. Each convolution is paired with a \ac{GELU} activation and Batch Normalization. The fusion block is defined as follows:
\begin{equation}
\begin{aligned}
f_{\text{concat }}&=\operatorname{Concat}\left(\operatorname{UpConv}\left(f_{\mathcal{E}}\right),\operatorname{UpConv}\left(f_{\mathcal{D}}\right)\right) \\
f_{\text{concat}}^{\prime}&=\phi\left(\mathrm{GConv}_{3\times3}\left(f_{\text {concat}},\operatorname{Group}=2\right)\right) \\
f_{\text{fusion}}^{\prime}&=\phi\left(\mathrm{PWConv}\left(f_{\text {concat}}^{\prime}\right)\right) \\
f_{\text{fusion}}&=\phi\left(\mathrm{PWConv}\left(f_{\text{fusion }}^{\prime}\right)\right)
\end{aligned}
\end{equation}
where $f_{\text {fusion}}$ represents the output fusion feature map from Fusion block, $\mathrm{GConv}$ is the group convolution layer, $f_{\mathcal{E}}$ and $f_{\mathcal{D}}$ represent the encoder and decoder features, respectively.

\subsection{Objective Function}

The overall training objective function is a joint loss with two parts: a \ac{Dice} loss $\mathcal{L}_{\text {Dice}}$, and a \ac{CE} loss $\mathcal{L}_{\text {CE}}$.
Mathematically, the joint loss can be expressed as:
$
\label{loss}
\mathcal{L}_{\text {total}}=\mathcal{L}_{\text {Dice}}\left(f_{\theta}\left(\boldsymbol{X}\right), \boldsymbol{Y}_{\mathrm{gt}}\right) + \mathcal{L}_{\text {CE}}\left(f_{\theta}\left(\boldsymbol{X}\right), \boldsymbol{Y}_{\mathrm{gt}}\right)
$
where $f_{\theta}\left(\boldsymbol{X}\right)$ is the output from LANCANet.




\section{Experiments}
\subsection{Datasets}

To evaluate the robustness and generalization of LANCANet, we conducted experiments on four ultrasound datasets: HC18, CCA, PSFHS, KEN-FH, and AFR-FH, covering fetal head segmentation, carotid artery segmentation, and multi-class pubic symphysis--fetal head segmentation. HC18, CCA, and PSFHS were randomly divided into training, validation, and testing sets using a 60\%/10\%/30\% split, while KEN-FH and AFR-FH were used as external test sets to assess cross-domain generalization. All images were converted to three-channel inputs and resized to $448 \times 448$ pixels for training and evaluation.

The HC18, CCA, PSFHS, and AFR-FH datasets are publicly available and were collected under the ethical approvals reported in their original publications. The KEN-FH dataset was collected with appropriate institutional ethical approval and informed consent from participants. All datasets used in this study were de-identified prior to analysis.

\subsubsection{HC18} Sourced from a database in the Netherlands, the HC18 dataset~\cite{Heuvel:2018_b} comprises standardized ultrasound images of fetuses' heads without any growth abnormalities. All scans were acquired from healthy singleton pregnancies. The images are captured by experienced sonographers. In each image, the sonographer draws an ellipse as \ac{ROI} to best fit the circumference of the fetal head. We utilized 999 annotated images with resolution $800 \times 540$ that are publicly accessible. The training set has 650 images; the validation set has 50 images; and the test set has 249 images.

\subsubsection{CCA} The \ac{CCA} ultrasound dataset~\cite{Momot:2022} consists of an aggregate of 1100 ultrasound images taken from 11 different subjects. Each subject was examined on both the left and right sides, resulting in a total of 100 images per subject. Each image is in a format of $3 \times 709 \times 749$. In this study, we used 690 images for training, 80 images for validation, and 330 images for testing. 

\subsubsection{PSFHS} The \ac{PS}-\ac{FH} dataset~\cite{Lu:2022,Bai:2025} contains 4,000 ultrasound images. It is designed for multi-class segmentation tasks, targeting two distinct classes. ultrasound labels within the set denote pixels as either 0, 1, or 2. Herein, 0 represents the background, 1 represents the \ac{PS}, and 2 represents the \ac{FH}. We selected 3,743 images that present both pubic symphysis and fetal head in this study. In particular, there are 2,930 images for training, 290 images for validation, and 523 images for testing.

\subsubsection{KEN-FH} To evaluate model generalization under domain shift, we collected an external fetal head ultrasound dataset (KEN-FH) from Nairobi, Kenya. It contains 57 scans from 24 patients across the second and third trimesters. All images were annotated by four experienced sonographers ($> 10$ years of experience). This dataset was acquired using six ultrasound systems from GE Healthcare and Mindray, providing diverse clinical settings for cross-device and cross-population evaluation. In this study, this dataset is only used for evaluating the generalization of the proposed methods under domain shift.

\subsubsection{AFR-FH} This African dataset~\cite{balcells:2023,sendra_balcells:2023} comprises five cohorts collected from clinical centers in Malawi, Egypt, Uganda, Ghana, and Algeria between November 2021 and February 2022. Each cohort contains 25 ultrasound scans from 25 pregnant women acquired using different ultrasound systems and imaging protocols across the second and/or third trimesters. The diversity of imaging devices and clinical environments makes it suitable for evaluating model generalization under domain shift.

\subsection{Implementation Details}

We set the number of training epochs to 200 with a batch size of 5 using stochastic gradient descent with a learning rate of 0.001, momentum of 0.9, and weight decay of $1e^{-4}$. For LANCANet, the {\it Token Adapter} performs $T=2$ refinement iterations by default unless otherwise specified in the ablation study. Our implementation was developed using Python 3.11.5, PyTorch 2.1.2, and CUDA 12.2. Training was conducted on a single NVIDIA 4090D GPU. Models were evaluated on the validation set after each epoch, and the best-performing weights were saved. A unified training protocol was adopted across all methods to ensure a controlled, reproducible, and consistent comparison while reducing potential bias introduced by extensive method-specific hyperparameter tuning. The selected optimization settings provided stable convergence across all evaluated methods. Model testing and computational cost evaluation were performed on an AMD EPYC 9654P CPU.

\subsection{Data Augmentation} 

During the training phase, we only applied data augmentations to the training data. The augmentation techniques include rotation within the range $(-20^\circ, 20^\circ)$ with probability $\mathcal{P}(\cdot)=0.5$, random brightness contrast with $\mathcal{P}(\cdot)=0.5$, random blur with probability $\mathcal{P}(\cdot)=0.3$, and Gaussian noise with probability $\mathcal{P}(\cdot)=0.3$. The detailed parameters of the data augmentation techniques are in our code repository on \href{https://anonymous.4open.science/r/LANCAN-21A0/README.md}{GitHub}.

\subsection{Comparative Analysis With SOTA Methods}

To validate the effectiveness of the proposed  LiteAdaNCA-Net, we performed a comparative analysis against a baseline method (UNet~\cite{Ronneberger:2015}) and several \ac{SOTA} lightweight methods: MobileNet V2~\cite{Sandler:2018}, SegFormer~\cite{Xie:2021}, UNeXt~\cite{Valanarasu:2022}, EGE-UNet~\cite{Ruan:2023}, LB-Unet~\cite{Xu:2024}, CMUNeXt~\cite{Tang:2024}, MK-Unet~\cite{Rahman:2025}, and PMFSNet~\cite{Zhong:2025}) on the HC18, \ac{CCA} and PSFHS datasets. Because HC18 and \ac{CCA} contain fewer than 1000 training images, SegFormer was initialized with publicly available ImageNet weights~\cite{Deng:2009} to improve optimization stability. All other models were trained from scratch.

\subsection{Evaluation Metrics} 
To quantitatively evaluate the performance of our proposed method, we use these metrics: \ac{DSC}, Jaccard, \ac{ASD}, and \ac{HD95}.
\[
\begin{split}
DSC &=\frac{2 \times T P}{F P+2 \times T P+F N} \\
Jaccard &=\frac{T P}{F P+T P+F N} \\
HD95 &=\max_{k95\%}{[d(X, Y), d(Y, X)]} \\ 
ASD &=\frac{d(X, Y)+d(Y, X)}{2} \\
\end{split}
\]

\section{Results}

\subsection{Quantitative Results}

\begin{table*}
\centering
\caption{The quantitative results on the HC18 and \ac{CCA} dataset. The best results are in {\bf bold}. The $2^\text{nd}$ best results are in \underline{underline}. 
}
\label{hc18_res}
\setlength{\tabcolsep}{3.8pt}
\begin{tabular}{l|cccc|cccc}
\hline
\multirow{2}{*}{\bf Method} & \multicolumn{4}{c}{\bf HC18} & \multicolumn{4}{|c}{\bf \ac{CCA}} \\
& DSC $\uparrow$ & Jaccard $\uparrow$ & \ac{HD95} $\downarrow$ & ASD $\downarrow$ & DSC $\uparrow$ & Jaccard $\uparrow$ & \ac{HD95} $\downarrow$ & ASD $\downarrow$ \\
\hline
UNet~\cite{Ronneberger:2015} & 93.64 (8.58) & 89.03 (12.42) & 39.32 (44.18) & 13.88 (17.13) & \underline{91.30 (7.69)} & \underline{84.70 (10.31) } & 30.06 (55.12) & 9.70 (17.79) \\
MobileNet~\cite{Sandler:2018} & 93.92 (7.41) & 89.30 (10.83) & 25.63 (33.55) & 9.18 (11.84) & 81.63 (15.14) & 71.04 (16.55) & 20.49 (13.82) & 8.32 (5.39) \\
SegFormer~\cite{Xie:2021} & \underline{96.40 (3.59)} & \underline{93.24 (5.41)} & \underline{14.63 (18.94)} & \underline{5.26 (5.30)} & 89.77 (6.22) & 81.94 (9.02) & 20.35 (26.14) & 7.03 (8.09) \\
UNeXt~\cite{Valanarasu:2022}  & 94.53 (5.78) & 90.11 (8.84) & 31.90 (35.72) & 10.77 (11.62) & 81.69 (13.89) & 70.79 (15.13) & 30.05 (29.46) & 12.02 (7.92) \\
EGE-UNet~\cite{Ruan:2023} & 94.63 (7.63) & 90.55 (10.30) & 16.06 (17.01) & 6.40 (6.50) & 86.98 (11.44) & 78.30 (13.44) & \underline{13.69 (21.43)} & \underline{5.03 (6.53)} \\
LB-Unet~\cite{Xu:2024} & 92.12 (10.13) & 86.66 (13.66) & 2.94 (37.46) & 11.34 (13.55) & 86.27 (13.95) & 77.76 (15.90) & 19.02 (34.24) & 6.82 (10.49) \\
CMUNeXt~\cite{Tang:2024} & 94.73 (7.71) & 90.77 (10.63) & 28.06 (37.55) & 9.69 (12.76) & 80.46 (13.29) & 68.85 (14.08) & 29.71 (22.52) & 13.29 (5.30) \\
MK-Unet~\cite{Rahman:2025} & 95.48 (5.34) & 91.75 (7.99) & 20.37 (29.79) & 7.09 (10.27) & 83.86 (8.92) & 73.05 (11.15) & 33.13 (26.58) & 13.79 (6.53) \\
PMFSNet~\cite{Zhong:2025} & 94.14 (6.05) & 89.44 (8.97) & 34.54 (35.64) & 11.15 (10.30) & 82.28 (19.79) & 73.44 (21.38) & 21.26 (27.09) & 7.46 (8.88) \\
\bf LANCANet (Ours) & \textbf{96.62 (3.09)} & \textbf{93.62 (5.16)} & \textbf{14.41 (19.65)} & \textbf{5.22 (5.80)} & \textbf{92.86 (5.03)} & \textbf{87.00 (7.11)} & \textbf{9.71 (26.35)} & \textbf{3.64 (8.57)} \\
\hline
\end{tabular}
\end{table*}

\subsubsection{HC18 and CCA} As shown in Table~\ref{hc18_res}, LANCANet achieves the best performance across all evaluation metrics, demonstrating both improved overlap accuracy and boundary delineation capability. On the HC18 dataset, LANCANet achieves the highest DSC (96.62\%) and Jaccard (93.62\%), while obtaining the lowest \ac{HD95} (14.41) and ASD (5.22). Compared with recent lightweight architectures such as MK-Unet, CMUNeXt, and PMFSNet, the proposed method shows consistent improvements across both region-based and boundary-based metrics. On the CCA dataset, LANCANet similarly achieves the best overall performance with a DSC of 92.86\%, Jaccard of 87.00\%, and the lowest \ac{HD95} (9.71) and ASD (3.64). Notably, improvements in boundary-based metrics suggest that the proposed \ac{NCA}-based refinement mechanism effectively localizes more precise contours and reduces segmentation errors.


\begin{table*}
\centering
\caption{The quantitative results on the PSFHS dataset. The best results are in {\bf bold}. The $2^\text{nd}$ best results are \underline{underlined}. 
}
\label{psfhs_res}
\setlength{\tabcolsep}{3.8pt}
\begin{tabular}{l|cccc|cccc}
\hline
\multirow{2}{*}{\bf Method} & \multicolumn{4}{c|}{\bf \acf{PS}} & \multicolumn{4}{c}{\bf \acf{FH}} \\
& DSC $\uparrow$ & Jaccard $\uparrow$ & \ac{HD95} $\downarrow$ & ASD $\downarrow$ & DSC $\uparrow$ & Jaccard $\uparrow$ & \ac{HD95} $\downarrow$ & ASD $\downarrow$ \\
\hline
UNet~\cite{Ronneberger:2015} & 77.89 (16.04) & \underline{66.14 (18.19)} & 22.40 (30.51) & \underline{7.84 (8.25)} & 85.55 (9.12) & 75.69 (11.98) & 49.21 (27.42) & 18.00 (10.61) \\
MobileNet~\cite{Sandler:2018} & 72.72 (15.92) & 59.24 (17.08) & \underline{21.19 (17.68)} & 8.59 (7.05) & \textbf{88.67 (7.12)} & \textbf{80.28 (10.01)} & \textbf{31.50 (22.75)} & \textbf{11.64 (7.59)} \\
SegFormer~\cite{Xie:2021} & \underline{78.45 (12.10)} & 65.86 (13.52) & \textbf{19.36 (28.31)} & \textbf{7.44 (8.04)} & 86.32 (9.23) & 76.92 (12.23) & 42.17 (28.79) & 15.47 (11.27) \\
UNeXt~\cite{Valanarasu:2022} & 76.44 (13.42) & 63.53 (15.53) & 26.64 (39.22) & 9.55 (10.64) & \underline{88.62 (6.73)} & \underline{80.15 (9.70)} & 40.42 (31.52) & 14.65 (10.30)\\
EGE-UNet~\cite{Ruan:2023} & 72.09 (15.69) & 58.42 (17.02) & 22.42 (14.88) & 8.88 (6.09) & 86.83 (9.99) & 77.86 (13.06) & \underline{35.46 (26.39)} & \underline{13.63 (10.26)} \\
LB-Unet~\cite{Xu:2024} & 70.26 (15.48) & 56.07 (16.34) & 22.92 (18.80) & 9.06 (7.04) & 85.82 (7.80) & 75.91 (11.11) & 38.13 (23.74) & 15.23 (9.86) \\
CMUNeXt~\cite{Tang:2024} & 71.30 (16.07) & 57.71 (17.61) & 24.59 (27.53) & 9.16 (8.74) & 85.24 (10.03) & 75.37 (13.19) & 41.06 (26.37) & 16.07 (10.45) \\
MK-Unet~\cite{Rahman:2025} & 74.16 (15.75) & 61.03 (16.98) & 23.21 (27.94) & 8.63 (8.64) & 85.09 (10.43) & 75.28 (13.71) & 41.65 (27.91) & 15.25 (9.54) \\
PMFSNet~\cite{Zhong:2025} & 75.60 (16.74) & 63.26 (18.72) & 30.50 (45.06) & 10.22 (12.32) & 87.86 (7.25) & 79.01 (10.40) & 41.19 (26.05) & 15.23 (9.88) \\
\bf LANCANet (Ours) & \textbf{79.00 (12.38)} & \textbf{66.74 (14.39)} & 25.95 (46.24) & 8.44 (9.78) & 85.63 (11.25) & 76.26 (14.41) & 39.10 (29.31)& 14.79 (10.36) \\
\hline
\end{tabular}
\end{table*}

\subsubsection{\ac{PS} and \ac{FH}} Table~\ref{psfhs_res} presents the results on the PSFHS dataset, which contains pixel-level annotations for both the \acf{FH} and the \acf{PS} within the same intrapartum transperineal ultrasound images. Overall, LANCANet achieves the best performance in segmenting the anatomically smaller and more challenging \ac{PS} structure. Specifically, LANCANet achieves the highest \ac{DSC} (79.00\%) and Jaccard (66.74\%) on \ac{PS}, a clinically critical region for assessing fetal head descent and guiding delivery decisions. For \ac{FH} segmentation, although certain baselines achieve better results in region-based and boundary-based metrics, LANCANet maintains competitive performance (85.63\% DSC, 76.26\% Jaccard, 39.10 \ac{HD95}, and 14.79 ASD) while substantially improving \ac{PS} segmentation accuracy. These results demonstrate the effectiveness of LANCANet for ultrasound segmentation of both small and large anatomical structures. 
This balanced performance is clinically meaningful, as precise delineation of the \ac{PS}-\ac{FH} relationship is essential for objective labor assessment. Despite its lightweight design, LANCANet achieves top-tier average performance compared to other lightweight models. It demonstrates an effective balance between accuracy and efficiency, supporting its potential for real-time clinical development.

Notably, the ablation study (Table~\ref{psfhs_ablation_t}) indicates that an appropriate number of \ac{NCA} refinement iterations can further improve segmentation performance on the PSFHS dataset. This finding suggests that iterative local refinement helps capture complex anatomical structures, although excessive refinement may lead to diminishing returns.

\begin{figure*}
\centering
\includegraphics[width=\textwidth]{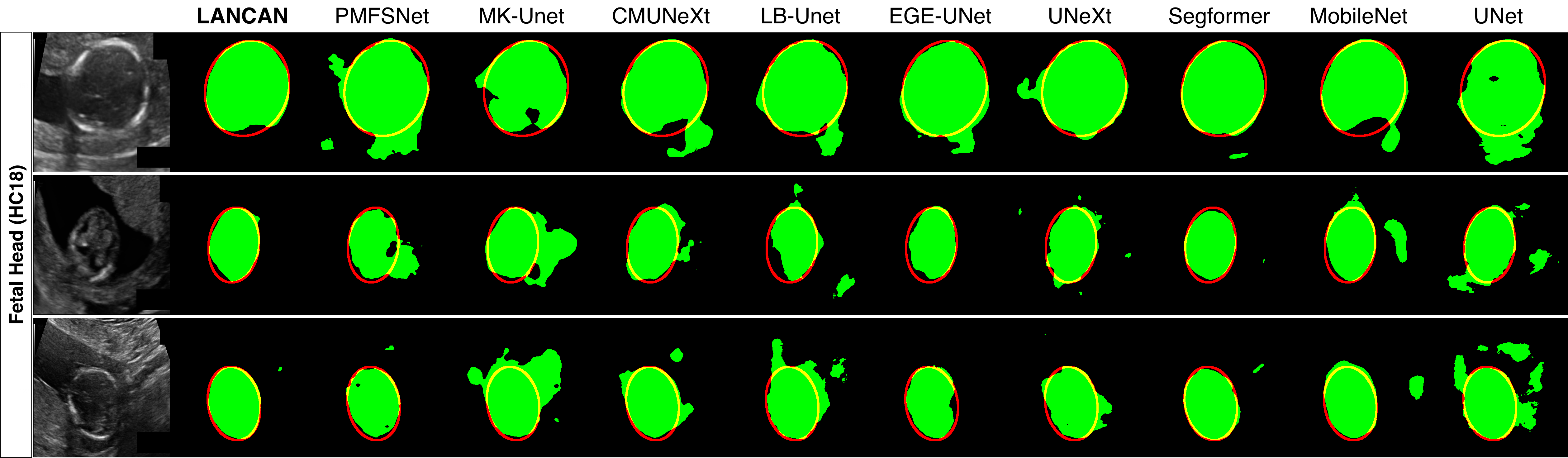}
\caption{Visual comparison of methods on HC18 dataset. The figure shows the \ac{GT} in red and the predicted results in green.}
\label{vis_res_hc18}
\end{figure*}

\begin{figure*}
\centering
\includegraphics[width=\textwidth]{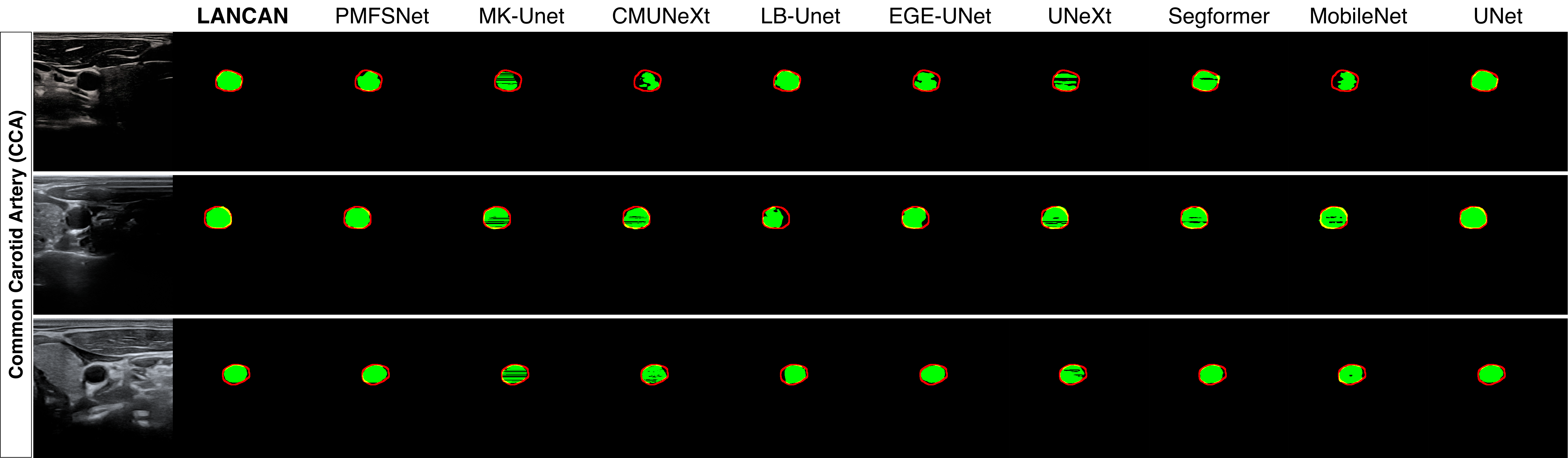}
\caption{Visual comparison of methods on CCA dataset. The figure shows the \ac{GT} in red and the predicted results in green.}
\label{vis_res_cca}
\end{figure*}

\begin{figure*}
\centering
\includegraphics[width=\textwidth]{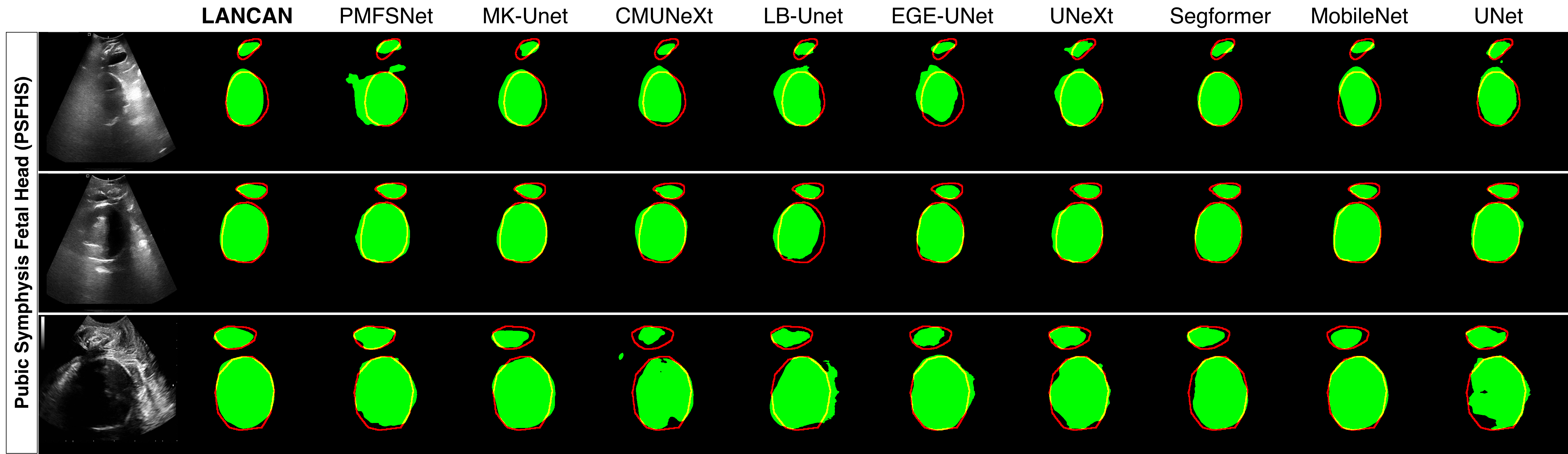}
\caption{Visual comparison of methods on PSFHS dataset. The figure shows the \ac{GT} in red and the predicted results in green.}
\label{vis_res_psfh}
\end{figure*}

\begin{figure*}
\centering
\includegraphics[width=\textwidth]{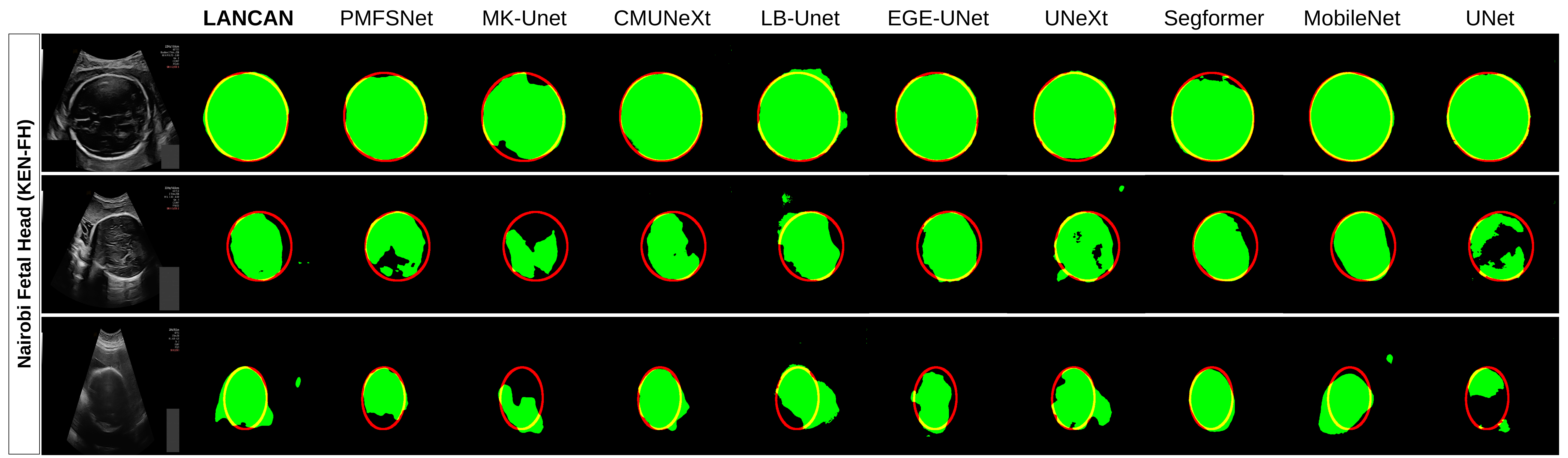}
\caption{Visual comparison of methods on KEN-FH dataset. The figure shows the \ac{GT} in red and the predicted results in green.}
\label{vis_res_kenfh}
\end{figure*}

\begin{figure*}
\centering
\includegraphics[width=\textwidth]{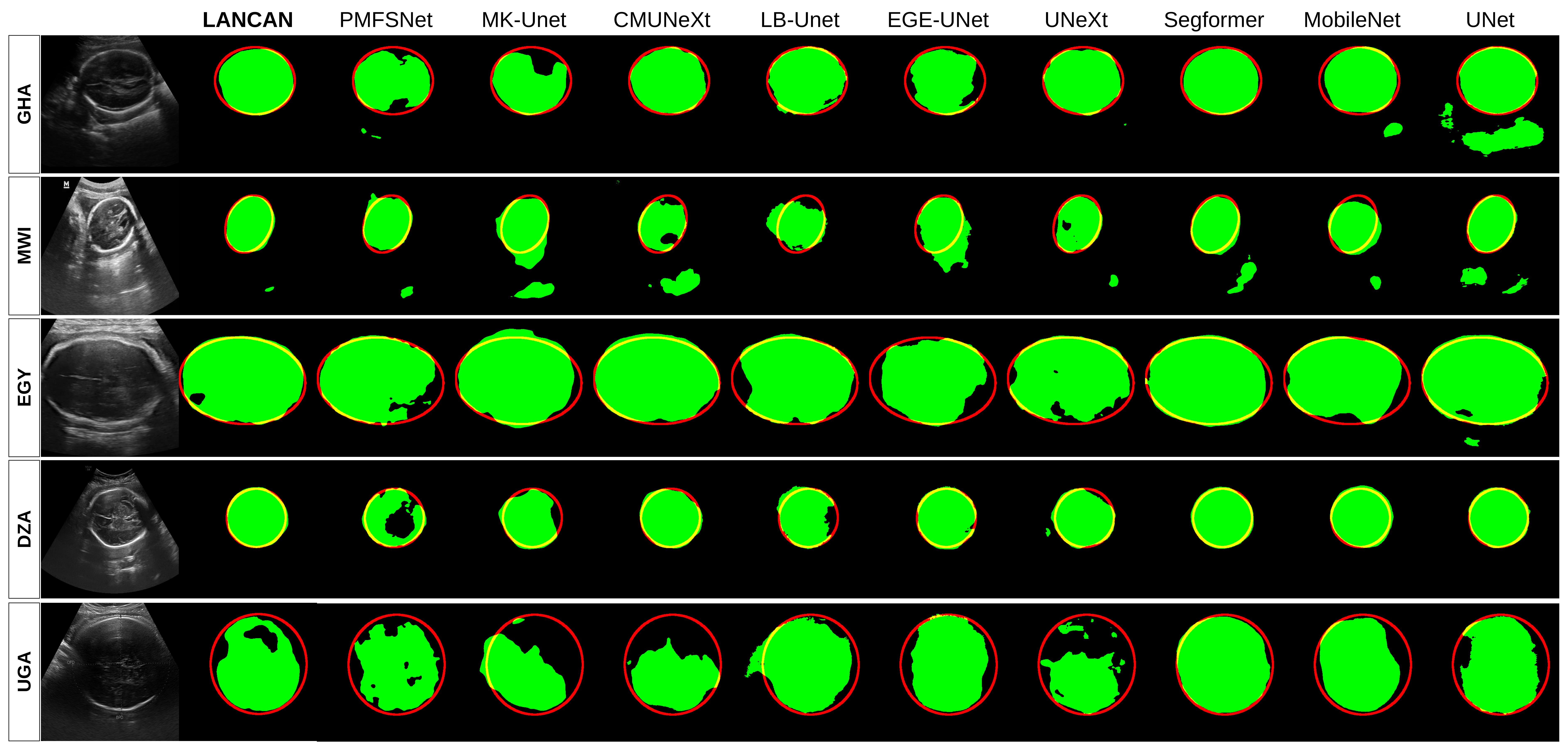}
\caption{Examples are shown from five African countries, GHA (Ghana), MWI (Malawi), EGY (Egypt), DZA (Algeria), and UGA (Uganda), illustrating the robustness of different methods under substantial cross-country and cross-device domain shifts. The \ac{GT} are in red and the predicted results are in green.}
\label{vis_res_afr_fh}
\end{figure*}


\subsection{External Validation under Domain Shift}


Table~\ref{ken_fg_res} presents the external evaluation results on two datasets collected from Africa. All models were trained on HC18 and directly evaluated on both datasets without fine-tuning. This experiment assesses model robustness to domain shifts arising from differences in population characteristics, ultrasound systems, acquisition protocols, and imaging conditions. 
On KEN-FH, SegFormer achieves the best DSC (92.90\%), Jaccard (87.01\%), HD95 (20.19), and ASD (7.79), while LANCANet obtains competitive performance with 89.63\% DSC, 81.91\% Jaccard, 32.92 HD95, and 12.31 ASD.
On AFR-FH, LANCANet achieves the second best \ac{DSC} (90.64\%) and Jaccard (83.82\%), whereas SegFormer attains the best performance in \ac{DSC} (93.01\%), Jaccard (87.22\%) and \ac{ASD} (13.91). Although LANCANet does not achieve the best score for all individual metrics on AFR-FH dataset, it consistently delivers competitive overlap- and boundary-based performance. 
Notably, SegFormer was initialized with ImageNet weights during training on HC18, which provides strong semantic representations under limited training data. Even without pre-training on ImageNet, our LANCANet model still maintains competitive performance in overlap- and boundary-based metrics.
This suggests that the proposed token-conditioned \ac{NCA} refinement effectively preserves structural consistency in unseen ultrasound domains. These results indicate the potential of LANCANet for more reliable deployment across diverse clinical settings.

\begin{table}
\centering
\caption{The quantitative results on two African fetal head datasets, KEN-\ac{FH} and AFR-\ac{FH}. The best results are in {\bf bold}. The $2^\text{nd}$ best results are \underline{underlined}.
}
\label{ken_fg_res}
\scalebox{0.9}{
\setlength{\tabcolsep}{2.5pt}
\begin{tabular}{l|cccc}
\hline
\bf Method & DSC $\uparrow$ & Jaccard $\uparrow$ & \ac{HD95} $\downarrow$ & ASD $\downarrow$\\
\hline
\rowcolor{tablegray} \bf KEN-\ac{FH} &  &  &  &  \\
UNet~\cite{Ronneberger:2015}         & 77.52(20.96)  & 67.49(24.83) & 55.61(37.65) & 18.79(11.69)  \\
MobileNet~\cite{Sandler:2018}        & 89.80(6.60)   & 82.10(10.31) & 45.38(40.41) & 15.45(11.86) \\
SegFormer~\cite{Xie:2021}            & \bf 92.90(4.21) & \bf 87.01(6.87) & \bf 20.19(20.96) & \bf 7.79(5.63) \\
UNeXt~\cite{Valanarasu:2022}         & 85.37(11.23)  & 76.02(15.86) & 55.73(37.52) & 19.46(11.60)  \\
EGE-UNet~\cite{Ruan:2023}            & \underline{90.47(7.20)} & \underline{83.24(9.91)} & \underline{23.06(14.94)} & \underline{9.67(6.46)} \\
LB-Unet~\cite{Xu:2024}               & 88.65(6.63)   & 80.20(9.89)  & 32.99(21.28) & 13.41(7.79)  \\
CMUNeXt~\cite{Tang:2024}             & 72.05(29.26)  & 63.34(31.09) & 50.31(35.28) & 16.84(10.10) \\
MK-Unet~\cite{Rahman:2025}           & 65.63(29.33)  & 54.92(28.26) & 64.40(42.08) & 20.87(11.05)  \\
PMFSNet~\cite{Zhong:2025}            & 85.16(10.88)  & 75.52(14.61) & 51.03(28.14) & 17.62(9.75)  \\
\bf LANCANet (Ours)                   & 89.63(7.33)   & 81.91(10.80) & 32.92(21.54) & 12.31(7.19) \\
\hline
\rowcolor{tablegray} \bf AFR-\ac{FH} &  &  &  &  \\
UNet~\cite{Ronneberger:2015} & 85.88(11.30) & 76.78(15.41) & 76.08(48.50) & 28.68(20.04) \\
MobileNet~\cite{Sandler:2018} & 88.59(8.46) & 80.44(12.12) & 47.16(40.00) & 17.20(13.81) \\
SegFormer~\cite{Xie:2021} & \bf 93.01(4.38) & \bf 87.22(6.91) & \underline{39.03(42.44)} & \bf 13.91(14.33) \\
UNeXt~\cite{Valanarasu:2022} & 84.65(12.74) & 75.25(17.11) & 66.48(41.25) & 24.18(15.55) \\
EGE-UNet~\cite{Ruan:2023} & 84.98(18.06) & 76.84(19.23) & \bf 38.68(28.54) & 15.26(10.77) \\
LB-Unet~\cite{Xu:2024} & 88.22(7.55) & 79.64(10.69) & 45.03(31.87) & 16.27(9.85) \\
CMUNeXt~\cite{Tang:2024} & 87.82(8.98) & 79.31(12.82) & 58.61(40.55) & 20.70(14.29) \\
MK-Unet~\cite{Rahman:2025} & 84.73(10.15) & 74.71(13.69) & 62.79(38.79) & 22.97(15.26) \\
PMFSNet~\cite{Zhong:2025} & 85.96(10.69) & 76.73(14.63) & 72.33(45.16) & 26.20(17.95) \\
\bf LANCANet (Ours) & \underline{90.64(8.69)} & \underline{83.82(11.90)} & 39.86(35.43) & \underline{14.23 (12.27)} \\
\hline
\end{tabular}
}
\end{table}

\subsection{Qualitative Results}

Fig.~\ref{vis_res_hc18}--\ref{vis_res_kenfh} present qualitative comparisons of segmentation results across the HC18, CCA, PSFHS, KEN-FH and AFR-FH datasets. We observe that LANCANet generates predictions that align more closely with the \ac{GT} boundaries. This indicates that LANCANet can generate smoother segmentation masks with fewer irregular regions and fewer isolated false predictions. These visual improvements are particularly evident in challenging cases with weak boundaries, low contrast, shadowing artifacts, and small anatomical structures. Moreover, LANCANet preserves anatomical shape more consistently. It generates contours that better match the expected elliptical structure of the fetal head, as illustrated in Fig.~\ref{vis_res_hc18} and Fig.~\ref{vis_res_kenfh}. 

The qualitative results on the AFR-FH dataset further demonstrate the robustness of LANCANet across ultrasound images acquired from five African countries (Ghana, Malawi, Egypt, Algeria, and Uganda). In Fig.~\ref{vis_res_afr_fh}, LANCANet consistently produces complete and anatomically plausible fetal head contours despite substantial variations in image quality, contrast, shadowing, and scanning devices. In contrast, MK-Unet and EGE-UNet show irregular boundary delineation, PMFSNet, CMUNeXt, and UNeXt tend to produce incomplete fetal head contours, and MobileNet and UNet generate isolated false-positive predictions outside the target anatomy. LANCANet better preserves the overall elliptical shape of the fetal head across different populations and imaging conditions.

These qualitative results demonstrate that LANCANet generalizes well across diverse African populations and ultrasound devices, supporting its suitability for fetal head segmentation in resource-constrained clinical settings.

\subsection{Statistic Analysis}

Table~\ref{stat_dsc} shows the statistical significance analysis of segmentation performance between LANCANet and competing methods across the HC18, CCA, PSFHS, KEN-FH and AFR-FH datasets using paired \emph{t}-tests. A significance threshold of $p < 0.05$ was adopted, while non-significant results ($p \geqslant 0.05$) are highlighted in red. On HC18 and CCA, LANCANet achieves significant performance gains against nearly all \ac{SOTA} models, with most $p$-values below 0.01. 
On the remaining datasets, several comparisons with recent \ac{SOTA} architectures, including SegFormer, EGE-UNet, LB-Unet, and MK-Unet, are not statistically significant.
Specifically, non-significant differences are observed between LANCANet and SegFormer on HC18 ($p=0.46$), PSFHS (\ac{PS} ($p=0.51$) and \ac{FH} ($p=0.33$)), as well as between LANCANet and MobileNet ($p=0.47$, $p=0.06$), EGE-UNet ($p=0.89$), and LB-Unet ($p=0.09$) on selected external dataset KEN-FH. Additionally, on the AFR-FH dataset, LANCANet shows statistically significant differences compared with all other methods except MobileNet ($p=0.06$).
These findings indicate that LANCANet achieves performance comparable to several strong recent architectures while maintaining a lightweight design. The statistical analysis shows that the proposed LANCANet achieves stable and competitive segmentation performance across multiple ultrasound datasets, with significant improvements over most \ac{SOTA} lightweight models.

\begin{table}
\centering
\caption{The significance $p$ of paired t-test between LANCANet and other methods on the overall \ac{DSC} across four datasets. *: $p < 0.05$. **: $p < 0.01$.
}
\label{stat_dsc}
\setlength{\tabcolsep}{4.2pt}
\begin{tabular}{l|cccccc}
\hline
\bf Versus & \bf HC18 & \bf CCA & \bf \ac{PS} & \bf \ac{FH} & \bf KEN-\ac{FH} & \bf AFR-\ac{FH} \\
\hline
\rowcolor{tablegray} LANCANet & & & & & & \\
\quad vs. UNet & ** & ** & 0.26 & 0.91        & ** & ** \\
\quad vs. MobileNet & ** & ** & ** & **       & 0.47 & 0.06 \\
\quad vs. SegFormer & 0.46 & ** & 0.51 & 0.33 & * & ** \\
\quad vs. UNeXt  & ** & ** & ** & **          & ** & **\\
\quad vs. EGE-UNet & ** & ** & ** & 0.10      & 0.89 & **\\
\quad vs. LB-Unet & ** & ** & ** & 0.78       & 0.09 & * \\
\quad vs. CMUNeXt & ** & ** & ** & 0.59       & ** & * \\
\quad vs. MK-Unet & ** & ** & ** & 0.47       & ** & ** \\
\quad vs. PMFSNet & ** & ** & ** & **         & ** & ** \\
\hline
\end{tabular}
\end{table}

\subsection{Computation Cost Comparison}

\begin{figure}
\centering
\includegraphics[width=\linewidth]{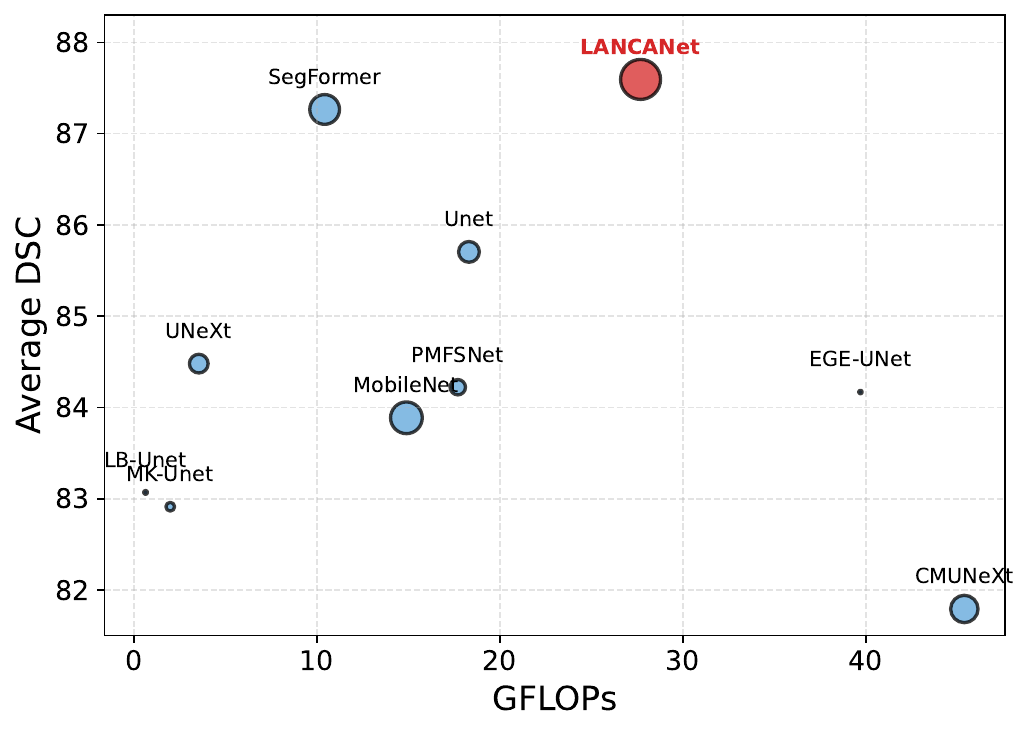}
\caption{Accuracy--efficiency comparison of different segmentation methods. Bubble size represents the number of model parameters.}
\label{acc_vs_complex}
\end{figure}

Table~\ref{comp_cost} presents the comparison of computational efficiency and overall segmentation performance across different methods. The reported \ac{DSC} values are averaged over the three ultrasound datasets, while computational efficiency is measured using the number of parameters, GFLOPs, and inference speed (FPS). Overall, LANCANet achieves the highest overall \ac{DSC} of 87.59\% and second best \ac{HD95} of 25.75. Fig.~\ref{acc_vs_complex} visualizes the trade-off between segmentation performance and computational complexity. Although LANCANet is not the smallest model, it achieves the highest average DSC while maintaining reasonable computational complexity. Compared with SegFormer, LANCANet achieves improved segmentation accuracy with a comparable model size, demonstrating a favorable balance between performance and computational cost. 

In Table~\ref{stat_overall_dsc}, statistical analysis shows that the performance improvement is significant for most methods ($p < 0.05$). However, the improvement over SegFormer was not statistically significant ($p \geq 0.05$), indicating comparable segmentation performance.

From the efficiency perspective, LANCANet requires only $6.70M$ parameters and 27.69 GFLOPs, remaining a compact and computationally efficient model among lightweight deep learning networks. Although LANCANet introduces higher latency than lightweight baselines such as UNeXt~\cite{Valanarasu:2022}, EGE-UNet~\cite{Ruan:2023} and LB-Unet~\cite{Xu:2024}, the additional computation leads to improved boundary-aware segmentation performance and overall accuracy. These results suggest that LANCANet provides a favorable balance between segmentation performance and computational efficiency for ultrasound imaging applications.


\begin{table}
\centering
\caption{Comparison of overall segmentation performance and computational cost across different methods on CPU. The $2^\text{nd}$ best results are \underline{underlined}.}
\label{comp_cost}
\scalebox{0.83}{
\begin{tabular}{l|ccc|cc}
\hline
\multirow{2}{*}{\bf Method} & \multicolumn{3}{c|}{\bf Efficiency} & \multicolumn{2}{c}{\bf Performance} \\
 & Param(M) $\downarrow$ &  GFLOPs $\downarrow$ & FPS $\uparrow$ & DSC $\uparrow$ & \ac{HD95} $\downarrow$ \\
\hline
UNet & 1.81 & 18.31 & 33.26 & 85.70(12.28) & 39.00(21.26) \\
MobileNet & 4.22 & 14.89 & 22.26 & 83.89(10.11) & 28.68(21.00)\\
SegFormer & 3.71 & 10.42 & 12.15 & \underline{87.26(6.62)} & 26.51(15.76)\\
UNeXt & 1.47 & 3.54 & \underline{33.85} & 84.48(10.63) & 35.91(19.77)\\
EGE-UNet & \bf 0.05 & 39.71 & 25.19 & 84.17(11.67) & \bf 24.37(14.71)\\
LB-Unet & \underline{0.06} & \bf 0.63 & \bf 38.40 & 83.07(10.26) & 51.52(16.75)\\
CMUNeXt & 3.15 & 45.39 & 8.09 & 81.79(14.22) & 34.50(20.49)\\
MK-Unet & 0.32 & \underline{1.98} & 16.63 & 82.85(13.32) & 34.23(20.22)\\
PMFSNet & 0.99 & 17.70 & 11.48 & 84.22(11.90) & 36.85(20.52)\\
LANCANet & 6.70 & 27.69 & 3.49 & \bf 87.59(7.96) & \underline{25.75(14.79)}\\
\hline
\end{tabular}
}
\end{table}

\begin{table}
\centering
\caption{The significance $p$ of paired t-test between LANCANet and other methods on the overall average \ac{DSC} across HC18, CCA, PSFHS, KEN-FH, and AFR-FH. *: $p < 0.05$. **: $p < 0.01$.
}
\label{stat_overall_dsc}
\begin{tabular}{l|cc}
\hline
\bf Versus & \bf DSC & \bf HD95 \\
\hline
\rowcolor{tablegray} LANCANet &  &   \\
\quad vs. UNet & ** & ** \\
\quad vs. MobileNet & ** & ** \\
\quad vs. SegFormer & 0.39 & 0.24 \\
\quad vs. UNeXt  & ** & ** \\
\quad vs. EGE-UNet & ** & * \\
\quad vs. LB-Unet & ** & ** \\
\quad vs. CMUNeXt & ** & ** \\
\quad vs. MK-Unet & ** & ** \\
\quad vs. PMFSNet & ** & ** \\
\hline
\end{tabular}
\end{table}

\subsection{Ablation Studies}

\begin{table*}
\centering
\caption{Ablation study on effectiveness of each component on the HC18 and \ac{CCA} dataset. The best results are in {\bf bold}. TGB: Token Guided Block.}
\label{ablation}
\setlength{\tabcolsep}{4pt}
\begin{tabular}{l|cccc|cccc}
\hline
\multirow{2}{*}{\bf Method} & \multicolumn{4}{c|}{\bf HC18} & \multicolumn{4}{c}{\bf \ac{CCA}} \\
& DSC $\uparrow$ & Jaccard $\uparrow$ & \ac{HD95} $\downarrow$ & ASD $\downarrow$ & DSC $\uparrow$ & Jaccard $\uparrow$ & \ac{HD95} $\downarrow$ & ASD $\downarrow$ \\
\hline
\bf LANCANet (Ours) & \textbf{96.62 (3.09)} & \textbf{93.62 (5.16)} & \textbf{14.41 (19.65)} & \textbf{5.22 (5.80)} & \textbf{92.86 (5.03)} & \textbf{87.00 (7.11)} & \textbf{9.71 (26.35)} & \textbf{3.64 (8.57)} \\
- w/o Token Adapter & 95.35 (5.43) & 91.54 (8.23) & 21.71 (28.55) & 7.55 (9.48) & 90.76 (11.33) & 84.39 (12.92) & 10.18 (17.83) & 3.73 (5.47) \\
- w/o Lite Transformer & 95.76 (5.26)  & 92.28 (8.12) & 20.86 (32.41) & 7.28 (11.46) & 92.66 (5.26) & 86.69 (7.62) & 9.81 (24.67) & 3.74 (9.34) \\
- w/o NCA TGB & 95.60 (4.76) & 91.97 (7.59) & 23.10 (32.07) & 7.82 (9.63) & 91.15 (5.36) & 84.13 (8.08) & 29.01 (61.87) & 9.50 (20.00) \\
\hline
\end{tabular}
\end{table*}

\begin{table*}
\centering
\caption{Ablation study on the influence of the \ac{NCA} update step $T$. The quantitative results on the PSFHS dataset. The best results are in {\bf bold}.}
\label{psfhs_ablation_t}
\setlength{\tabcolsep}{5pt}
\begin{tabular}{l|cc|cc|cc|cccc}
\hline
\multirow{2}{*}{\bf NCA Step} & \multicolumn{2}{c|}{\bf \acf{PS}} & \multicolumn{2}{c|}{\bf \acf{FH}} & \multicolumn{2}{c|}{\bf Avg.} & \bf Params $\downarrow$ & \multirow{2}{*}{\bf GFLOPs $\downarrow$} & \multirow{2}{*}{\bf FPS $\uparrow$} & \bf Latency $\downarrow$ \\
& DSC $\uparrow$ & \ac{HD95} $\downarrow$ & DSC $\uparrow$ & \ac{HD95} $\downarrow$ & DSC $\uparrow$ & \ac{HD95} $\downarrow$ & (M) &  &  & (ms)\\
\hline
$T=1$ & 77.87 (11.25) & 23.48 (37.27) & 89.65 (6.95) & 36.42 (29.20) & 83.76 & 29.95 & 6.70 & 27.69 & 3.49 & \bf 251.49 \\
$T=2$ & \bf 79.00 (12.38) & 25.95 (46.24) & 85.63 (11.25) & 39.10 (29.31) & 82.32 & 32.52 & 6.70 & 27.69 & 3.49 & 285.79 \\
$T=4$ & 78.18 (11.94) & \bf 17.60 (16.93) & \bf 90.15 (7.40) & \bf 30.71 (23.43) & \bf 84.17 & \bf 24.16 & 6.70 & 27.69 & 3.64 & 286.34 \\
$T=8$ & 75.93 (14.05) & 19.89 (14.48) & 89.89 (7.64) & 32.49 (26.13) & 82.91 & 26.19 & 6.70 & 27.69 & 3.64 & 302.02 \\
\hline
\end{tabular}
\end{table*}

\subsubsection{Module Ablation Analysis} Table~\ref{ablation} presents an ablation study evaluating the contribution of each component on the HC18 and CCA datasets. Removing the {\it Token Adapter} leads to a notable decrease in performance across all metrics, demonstrating the critical role of \ac{NCA}-based refinement. In contrast, removing the Lite Transformer block results in a marginal performance drop, which can be attributed to the limited training data where global attention provides weak benefits. The substantial degradation observed when removing the \ac{NCA} Token Guided Block confirms that \ac{NCA}-based refinement is not auxiliary, but a necessary component for achieving the reported performance gains.

\subsubsection{NCA Step Analysis} We further investigate the influence of the \ac{NCA} refinement step $T$ on segmentation performance and computational efficiency, shown in Table~\ref{psfhs_ablation_t}. On the PSFHS dataset, increasing $T$ from 1 to 4 improves segmentation accuracy. When $T=4$, LANCANet achieves the highest average \ac{DSC} (84.17\%) and the lowest average \ac{HD95} (24.16), which are the best results for both \ac{PS} and \ac{FH} structures. This suggests that appropriate iterative refinement enables more effective propagation of local contextual information and improves boundary delineation. However, further increasing the refinement step $T$ to 8 results in a drop in segmentation performance and an increase in computational cost, indicating potential over-refinement of feature representations.

In terms of inference efficiency, varying $T$ does not affect the number of model parameters (6.70M) or GFLOPs (27.69). However, increasing $T$ results in higher inference latency and lower throughput. For instance, latency rises from 274.70 $\text{ms}$ at $T=4$ to 302.02 $\text{ms}$ at $T=8$. Therefore, $T=4$ offers the best balance between segmentation performance and inference efficiency on the PSFHS dataset.

\section{Discussion}

In this study, we introduced LANCANet, a lightweight model with \acf{NCA}-based adapters for ultrasound image segmentation. We also investigated whether \ac{NCA}-based adapters can improve segmentation performance while maintaining computational efficiency. Experimental results across multiple ultrasound datasets show that LANCANet achieves a balance between segmentation accuracy, model complexity, and generalization capability. In particular, LANCANet consistently achieved competitive \ac{DSC} under different clinical settings. These findings suggest that \ac{NCA}-based adapters offer an effective strategy for developing efficient and robust ultrasound segmentation models, with potential advantages for deployment in resource-constrained settings.

The observed performance of LANCANet may be attributed to the proposed \ac{NCA}-based adapters. Unlike \ac{CNN} encoders that directly transform feature representations into segmentation outputs, LANCANet introduces an iterative cellular refinement process that updates local representations while incorporating global structural information progressively. Specifically, the {\it Structure Token Extraction} module compresses cellular states into compact global descriptors to capture image-level structural context. The {\it Token FiLM} module then incorporates this contextual information back into the cellular states through feature-wise modulation. This step allows local updates to be guided by global anatomical structures. We hypothesize that integrating global context with local refinement is beneficial for ultrasound imaging, where low contrast, speckle noise, and ambiguous boundaries often lead to inaccurate segmentation. Furthermore, the proposed design enables LANCANet to be a lightweight model suitable for efficient deployment.

As a lightweight model, LANCANet may provide practical value for ultrasound applications in resource-constrained settings, where computational infrastructure is often limited. 
SegFormer achieved the highest performance on KEN-FH and AFR-FH, likely because it was initialized with pretrained ImageNet weights. 
Nevertheless, LANCANet was trained from scratch and maintained competitive performance across both KEN-FH and AFR-FH datasets, suggesting that token-conditioned \ac{NCA} refinement enhances feature robustness without external pretraining.
Therefore, LANCANet demonstrates the potential for deployment on edge devices to support fetal head segmentation across different imaging devices and populations in resource-constrained settings.

This study has a few limitations. First, all models were trained under a unified training protocol without individual hyperparameter tuning to ensure controlled, reproducible, and fair comparison. However, further performance improvements may be achieved through model optimization. Second, this work focuses on segmentation tasks, while other ultrasound tasks (e.g., classification, detection, and regression) remain unexplored. Future work will extend the evaluation to additional anatomical structures, such as cardiac and breast ultrasound, to further evaluate model generalization. We also plan to investigate deployment on edge devices and evaluate real-time inference performance to better support practical clinical applications.

\section{Conclusion}
This paper presented LANCANet, a lightweight ultrasound segmentation framework that integrates token-conditioned \acf{NCA} adapters for iterative feature refinement. By combining global structural guidance with local cellular updates, LANCANet improves boundary delineation while maintaining a favorable balance between segmentation accuracy and computational efficiency. Comprehensive experiments on multiple ultrasound datasets, including external validation under domain shift, show that LANCANet achieves competitive or superior performance compared with recent lightweight CNN- and transformer-based methods. These findings indicate that token-conditioned \ac{NCA} refinement is an effective strategy for robust and efficient ultrasound segmentation. Owing to its robustness across different populations and imaging devices, LANCANet provides a practical solution for \acf{AI}-assisted ultrasound analysis in \acf{POC} and resource-constrained settings.

\section*{Acknowledgment}

This work was funded by Taighde \'{E}ireann – Research Ireland through the Research Ireland Centre for Research Training in Machine Learning (18/CRT/6183). This research was supported by a grant from the European Research Council (ERC) under the European Union's Horizon Europe research and innovation programme (AIMIX project - Grant Agreement No. 101044779).

\section*{Disclosure of Interests}
The authors have no competing interests to declare that are relevant to the content of this article.

\section*{References}

\bibliographystyle{IEEEbib}
\bibliography{ref}

@article{Heuvel:2018_b,
    author = {van den Heuvel, Thomas L. A. AND de Bruijn, Dagmar AND de Korte, Chris L. AND Ginneken, Bram van},
    journal = {PLOS ONE},
    publisher = {Public Library of Science},
    title = {Automated measurement of fetal head circumference using 2D ultrasound images},
    year = {2018},
    month = Aug,
    volume = {13},
    pages = {1-20},
    number = {8},
}

@article{Dosovitskiy:2020_b,
    title={An Image is Worth 16x16 Words: Transformers for Image Recognition at Scale},
    author={Dosovitskiy, Alexey and Beyer, Lucas and Kolesnikov, Alexander and Weissenborn, Dirk and Zhai, Xiaohua and Unterthiner, Thomas and et al.},
    journal={ICLR},
    year={2021}
}

@article{Salomon:2011,
    author = {Salomon, L. J. and Alfirevic, Z. and Berghella, V. and Bilardo, C. and Hernandez-Andrade, E. and Johnsen, S. L. and Kalache, K. and Leung, K.-Y. and Malinger, G. and Munoz, H. and Prefumo, F. and Toi, A. and Lee, W. and on behalf of the ISUOG Clinical Standards Committee},
    title = {Practice guidelines for performance of the routine mid-trimester fetal ultrasound scan},
    journal = {Ultrasound in Obstetrics \& Gynecology},
    volume = {37},
    number = {1},
    pages = {116-126},
    year = {2011}
}

@InProceedings{Ronneberger:2015,
    title={{U-Net}: Convolutional Networks for Biomedical Image Segmentation}, 
    author={Ronneberger, Olaf and Fischer, Philipp and Brox, Thomas},
    year={2015},
    volume={9351},
    booktitle="International Conference on Medical Image Computing and Computer-Assisted Intervention",
    publisher="Springer",
    pages="234--241",
    isbn="978-3-319-24574-4",
    eprint={1505.04597},
    archivePrefix={arXiv},
    primaryClass={cs.CV}
}

@InProceedings{Sandler:2018,
    author = {Sandler, Mark and Howard, Andrew and Zhu, Menglong and Zhmoginov, Andrey and Chen, Liang-Chieh},
    title = {MobileNetV2: Inverted Residuals and Linear Bottlenecks},
    booktitle = {Proceedings of the IEEE Conference on Computer Vision and Pattern Recognition (CVPR)},
    month = {June},
    year = {2018}
}

@inproceedings{Xie:2021,
    title={SegFormer: Simple and Efficient Design for Semantic Segmentation with Transformers},
    author={Xie, Enze and Wang, Wenhai and Yu, Zhiding and Anandkumar, Anima and Alvarez, Jose M and Luo, Ping},
    booktitle={Neural Information Processing Systems (NeurIPS)},
    year={2021}
}

@INPROCEEDINGS{Valanarasu:2022,
    title     = "{UNeXt}: {MLP-Based} Rapid Medical Image Segmentation Network",
    booktitle = "International Conference on Medical Image Computing and Computer-Assisted Intervention ",
    author    = "Valanarasu, Jeya Maria Jose and Patel, Vishal M",
    publisher = "Springer Nature Switzerland",
    pages     = "23--33",
    year      =  2022
}

@INPROCEEDINGS{Ruan:2023,
    title     = "{EGE-UNet}: An Efficient Group Enhanced {UNet} for Skin Lesion Segmentation",
    booktitle = "International Conference on Medical Image Computing and Computer-Assisted Intervention ",
    author    = "Ruan, Jiacheng and Xie, Mingye and Gao, Jingsheng and Liu, Ting and Fu, Yuzhuo",
    publisher = "Springer",
    pages     = "481--490",
    year      =  2023
}

@INPROCEEDINGS{Xu:2024,
    title     = "{LB-UNet}: A Lightweight {Boundary-Assisted} {UNet} for Skin Lesion Segmentation",
    booktitle = "International Conference on Medical Image Computing and Computer-Assisted Intervention ",
    author    = "Xu, Jiahao and Tong, Lyuyang",
    publisher = "Springer Nature Switzerland",
    pages     = "361--371",
    year      =  2024
}

@INPROCEEDINGS{Tang:2024,
    author={Tang, Fenghe and Ding, Jianrui and Quan, Quan and Wang, Lingtao and Ning, Chunping and Zhou, S. Kevin},
    booktitle={2024 IEEE International Symposium on Biomedical Imaging (ISBI)}, 
    title={CMUNEXT: An Efficient Medical Image Segmentation Network Based on Large Kernel and Skip Fusion}, 
    year={2024},
    volume={},
    number={},
    pages={1-5},
}

@inproceedings{Rahman:2025,
    title={Mk-unet: Multi-kernel lightweight cnn for medical image segmentation},
    author={Rahman, Md Mostafijur and Marculescu, Radu},
    booktitle={Proceedings of the IEEE/CVF International Conference on Computer Vision},
    pages={1042--1051},
    year={2025}
}

@article{Zhong:2025,
    title = {PMFSNet: Polarized multi-scale feature self-attention network for lightweight medical image segmentation},
    journal = {Computer Methods and Programs in Biomedicine},
    volume = {261},
    pages = {108611},
    year = {2025},
    issn = {0169-2607},
    author = {Jiahui Zhong and Wenhong Tian and Yuanlun Xie and Zhijia Liu and Jie Ou and Taoran Tian and Lei Zhang},
}

@article{Momot:2022,
    title     = "Common carotid artery ultrasound images",
    author    = "Agata Momot",
    journal = "Mendeley Data",
    doi = "10.17632/d4xt63mgjm.1",
    year      =  2022
}

@article{Bai:2025,
    title = {PSFHS challenge report: Pubic symphysis and fetal head segmentation from intrapartum ultrasound images},
    journal = {Medical Image Analysis},
    volume = {99},
    pages = {103353},
    year = {2025},
    issn = {1361-8415},
    author = {Jieyun Bai and Zihao Zhou and Zhanhong Ou and Gregor Koehler and Raphael Stock and Klaus Maier-Hein and Marawan Elbatel and et al.}
}

@article{Gilpin:2019,
    title={Cellular automata as convolutional neural networks},
    author={Gilpin, William},
    journal={Physical Review E},
    volume={100},
    number={3},
    pages={032402},
    year={2019},
    publisher={APS}
}

@article{Mordvintsev:2020,
    author = {Mordvintsev, Alexander and Randazzo, Ettore and Niklasson, Eyvind and Levin, Michael},
    title = {Growing Neural Cellular Automata},
    journal = {Distill},
    year = {2020},
}

@inproceedings{Wulff:1992,
    author = {Wulff, N. and Hertz, J A},
    booktitle = {Advances in Neural Information Processing Systems},
    editor = {S. Hanson and J. Cowan and C. Giles},
    pages = {},
    publisher = {Morgan-Kaufmann},
    title = {Learning Cellular Automaton Dynamics with Neural Networks},
    volume = {5},
    year = {1992}
}

@InProceedings{Yang:2026,
    author="Yang, Chen and Deutges, Michael and Liu, Jingsong and Li, Han and Navab, Nassir and Marr, Carsten and Sadafi, Ario",
    title="Attention Pooling Enhances NCA-Based Classification of Microscopy Images",
    booktitle="Machine Learning in Medical Imaging",
    year="2026",
    publisher="Springer Nature Switzerland",
    pages="583--593",
}

@inproceedings{Xu:2024b,
    author = {Xu, Yitao and Zhang, Tong and S\"{u}sstrunk, Sabine},
    title = {AdaNCA: neural cellular automata as adaptors for more robust vision transformer},
    year = {2024},
    isbn = {9798331314385},
    publisher = {Curran Associates Inc.},
    booktitle = {Proceedings of the 38th International Conference on Neural Information Processing Systems},
    articleno = {809},
    numpages = {38},
}

@InProceedings{Deutges:2026,
    author="Deutges, Michael and Yang, Chen and Salehi, Raheleh and Navab, Nassir and Marr, Carsten and Sadafi, Ario", 
    title="Neural Cellular Automata for Weakly Supervised Segmentation of White Blood Cells",
    booktitle="Efficient Medical Artificial Intelligence",
    year="2026",
    publisher="Springer Nature Switzerland",
    pages="289--298",
    isbn="978-3-032-13961-0"
}

@ARTICLE{Krumb:2025,
    title    = "{eNCApsulate}: neural cellular automata for precision diagnosis on capsule endoscopes",
    author   = "Krumb, Henry John and Mukhopadhyay, Anirban",
    journal  = "International Journal of Computer Assisted Radiology and Surgery",
    month    =  jul,
    year     =  2025
}

@inproceedings{Mittal:2025,
    title={Medsegdiffnca: Diffusion models with neural cellular automata for skin lesion segmentation},
    author={Mittal, Avni and Kalkhof, John and Mukhopadhyay, Anirban and Bhavsar, Arnav},
    booktitle={IEEE 38th International Symposium on Computer-Based Medical Systems},
    pages={35--40},
    year={2025},
    organization={IEEE}
}

@inproceedings{Kalkhof:2023,
    title={Med-NCA: Robust and Lightweight Segmentation with Neural Cellular Automata},
    author={Kalkhof, John and Gonz{\'a}lez, Camila and Mukhopadhyay, Anirban},
    booktitle={International Conference on Information Processing in Medical Imaging},
    pages={705--716},
    year={2023},
    organization={Springer}
}

@inproceedings{Kalkhof:2023b,
    title={M3D-NCA: Robust 3D Segmentation with Built-In Quality Control},
    author={Kalkhof, John and Mukhopadhyay, Anirban},
    booktitle={International Conference on Medical Image Computing and Computer-Assisted Intervention },
    pages={169--178},
    year={2023},
    organization={Springer}
}

@article{Zhou:2022,
    title = {LAEDNet: A Lightweight Attention Encoder–Decoder Network for ultrasound medical image segmentation},
    journal = {Computers and Electrical Engineering},
    volume = {99},
    pages = {107777},
    year = {2022},
    issn = {0045-7906},
    author = {Quan Zhou and Qianwen Wang and Yunchao Bao and Lingjun Kong and Xin Jin and Weihua Ou},
}

@article{Pietsch:2021,
    title = {APPLAUSE: Automatic Prediction of PLAcental health via U-net Segmentation and statistical Evaluation},
    journal = {Medical Image Analysis},
    volume = {72},
    pages = {102145},
    year = {2021},
    issn = {1361-8415},
    author = {Maximilian Pietsch and Alison Ho and Alessia Bardanzellu and Aya Mutaz Ahmad Zeidan and et al.},
}

@ARTICLE{Lee:2020,
    title    = "{Point-of-Care} Ultrasound",
    author   = "Lee, Linda and DeCara, Jeanne M",
    journal  = "Current Cardiology Reports",
    volume   =  22,
    number   =  11,
    pages    = "149",
    month    =  sep,
    year     =  2020
}

@Article{Self:2022,
    author="Self, Alice and Chen, Qingchao and Desiraju, Bapu Koundinya and Dhariwal, Sumeet and Gleed, Alexander D and Mishra, Divyanshu and et al.",
    title="Developing Clinical Artificial Intelligence for Obstetric Ultrasound to Improve Access in Underserved Regions: Protocol for a Computer-Assisted Low-Cost Point-of-Care UltraSound (CALOPUS) Study",
    journal="JMIR Res Protoc",
    year="2022",
    month="Sep",
    day="1",
    volume="11",
    number="9",
    pages="e37374",
    issn="1929-0748",
}

@ARTICLE{Zhou:2020,
    author={Zhou, Zixia and Wang, Yuanyuan and Guo, Yi and Qi, Yanxing and Yu, Jinhua},
    journal={IEEE Transactions on Biomedical Engineering}, 
    title={Image Quality Improvement of Hand-Held Ultrasound Devices With a Two-Stage Generative Adversarial Network}, 
    year={2020},
    volume={67},
    number={1},
    pages={298-311},
}

@ARTICLE{Noble:2006,
    author={Noble, J.A. and Boukerroui, D.},
    journal={IEEE Transactions on Medical Imaging}, 
    title={Ultrasound image segmentation: a survey}, 
    year={2006},
    volume={25},
    number={8},
    pages={987-1010},
}

@misc{Ferreira:2025,
    title={SAS: Segment Anything Small for Ultrasound--A Non-Generative Data Augmentation Technique for Robust Deep Learning in Ultrasound Imaging},
    author={Ferreira, Danielle L and Gangopadhyay, Ahana and Chang, Hsi-Ming and Soni, Ravi and Avinash, Gopal},
    eprint = {2503.05916},
    archivePrefix= {arXiv},
    primaryClass = {eess.IV},
    note = {arXiv:2503.05916},
    year={2025}
}

@article{Vaswani:2017,
    title={Attention is all you need},
    author={Vaswani, Ashish and Shazeer, Noam and Parmar, Niki and Uszkoreit, Jakob and Jones, Llion and Gomez, Aidan N and Kaiser, {\L}ukasz and Polosukhin, Illia},
    journal={Advances in neural information processing systems},
    volume={30},
    year={2017}
}

@article{Trockman:2023,
    title={Patches Are All You Need?},
    author={Asher Trockman and J Zico Kolter},
    journal={Transactions on Machine Learning Research},
    issn={2835-8856},
    year={2023}
}

@inproceedings{Tang:2023,
    title={Cmu-net: a strong convmixer-based medical ultrasound image segmentation network},
    author={Tang, Fenghe and Wang, Lingtao and Ning, Chunping and Xian, Min and Ding, Jianrui},
    booktitle={IEEE 20th international symposium on biomedical imaging (ISBI)},
    pages={1--5},
    year={2023},
}

@inproceedings{Perez:2018,
    author = {Perez, Ethan and Strub, Florian and de Vries, Harm and Dumoulin, Vincent and Courville, Aaron},
    title = {FiLM: visual reasoning with a general conditioning layer},
    year = {2018},
    isbn = {978-1-57735-800-8},
    booktitle = {Proceedings of the 32nd AAAI Conference on Artificial Intelligence},
    articleno = {483},
    numpages = {10},
    series = {AAAI-18}
}

@article{Lu:2022,
    title = {The JNU-IFM dataset for segmenting pubic symphysis-fetal head},
    journal = {Data in Brief},
    volume = {41},
    pages = {107904},
    year = {2022},
    issn = {2352-3409},
    doi = {10.1016/j.dib.2022.107904},
    author = {Yaosheng Lu and Mengqiang Zhou and Dengjiang Zhi and Minghong Zhou and Xiaosong Jiang and Ruiyu Qiu and Zhanhong Ou and Huijin Wang and Di Qiu and Mei Zhong and Xiaoxing Lu and Gaowen Chen and Jieyun Bai}
}

@ARTICLE{Boumeridja:2025,
    title    = "Enhancing fetal ultrasound image quality and anatomical plane recognition in low-resource settings using super-resolution models",
    author   = "Boumeridja, Hafida and Ammar, Mohammed and Alzubaidi, Mahmood and Mahmoudi, Sa{\"\i}d and Benamer, Lamya Nawal and Agus, Marco and Househ, Mowafa and Lekadir, Karim and El Habib Daho, Mostafa",
    journal  = "Scientific Reports",
    volume   =  15,
    number   =  1,
    pages    = "8376",
    month    =  mar,
    year     =  2025
}

@article{Liu:2020,
    title = {A survey on U-shaped networks in medical image segmentations},
    journal = {Neurocomputing},
    volume = {409},
    pages = {244-258},
    year = {2020},
    issn = {0925-2312},
    doi = {10.1016/j.neucom.2020.05.070},
    author = {Liangliang Liu and Jianhong Cheng and Quan Quan and Fang-Xiang Wu and Yu-Ping Wang and Jianxin Wang},
}

@article{Bian:2025,
    title = {ThreeF-Net: Fine-grained feature fusion network for breast ultrasound image segmentation},
    journal = {Computers in Biology and Medicine},
    volume = {194},
    pages = {110527},
    year = {2025},
    issn = {0010-4825},
    doi = {10.1016/j.compbiomed.2025.110527},
    author = {Xuesheng Bian and Jia Liu and Sen Xu and Weiquan Liu and Leyi Mei and Chaoshen Xiao and Fan Yang},
}

@ARTICLE{Isensee:2021,
    title    = "{nnU-Net}: a self-configuring method for deep learning-based biomedical image segmentation",
    author   = "Isensee, Fabian and Jaeger, Paul F and Kohl, Simon A A and Petersen, Jens and Maier-Hein, Klaus H",
    journal  = "Nature Methods",
    volume   =  18,
    number   =  2,
    pages    = "203--211",
    month    =  feb,
    year     =  2021
}

@misc{balcells:2023,
    author = {Carla Sendra-Balcells and Víctor M. Campello and Jordina Torrents-Barrena and Yahya Ali Ahmed and Mustafa Elattar and Benard Ohene-Botwe and Pempho Nyangulu and William Stones and Mohammed Ammar and Lamya Nawal Benamer and Harriet Nalubega Kisembo and Senai Goitom Sereke and Sikolia Z. Wanyonyi and Marleen Temmerman and Eduard Gratacós and Elisenda Bonet and Elisenda Eixarch and Kamil Mikolaj and Martin Grønnebæk Tolsgaard and Karim Lekadir},
    title = {Maternal fetal ultrasound planes from low-resource imaging settings in five African countries},
    month = feb,
    year = 2023,
    publisher = {Zenodo},
    version = {v1.0},
    doi = {10.5281/zenodo.7540448},
}

@article{sendra_balcells:2023,
    title = {Generalisability of fetal ultrasound deep learning models to low-resource imaging settings in five {African} countries},
    volume = {13},
    issn = {2045-2322},
    doi = {10.1038/s41598-023-29490-3},
    language = {en},
    number = {1},
    urldate = {2023-09-15},
    journal = {Scientific Reports},
    author = {Sendra-Balcells, Carla and Campello, Víctor M. and Torrents-Barrena, Jordina and Ahmed, Yahya Ali and Elattar, Mustafa and Ohene-Botwe, Benard and Nyangulu, Pempho and Stones, William and Ammar, Mohammed and Benamer, Lamya Nawal and Kisembo, Harriet Nalubega and Sereke, Senai Goitom and Wanyonyi, Sikolia Z. and Temmerman, Marleen and Gratacós, Eduard and Bonet, Elisenda and Eixarch, Elisenda and Mikolaj, Kamil and Tolsgaard, Martin Grønnebæk and Lekadir, Karim},
    month = feb,
    year = {2023},
    pages = {2728},
}

@INPROCEEDINGS{Deng:2009,
    author={Deng, Jia and Dong, Wei and Socher, Richard and Li, Li-Jia and Kai Li and Li Fei-Fei},
    booktitle={2009 IEEE Conference on Computer Vision and Pattern Recognition}, 
    title={ImageNet: A large-scale hierarchical image database}, 
    year={2009},
    volume={},
    number={},
    pages={248-255},
    doi={10.1109/CVPR.2009.5206848}
}

\end{document}